\documentclass[a4paper,12pt]{article}

\usepackage[left=3.5cm,right=3.5cm,top=3.5cm,bottom=3.5cm,twoside]{geometry}
\usepackage{amsfonts}
\usepackage{amsthm}
\usepackage{amsmath}
\usepackage{amssymb}
\usepackage{array}
\usepackage{bbding}
\usepackage{bbm}
\usepackage{caption}
\usepackage{colortbl} 
\usepackage{dsfont}
\usepackage{eurosym}
\usepackage{float}
\usepackage{graphicx}
\usepackage{hyphenat}
\usepackage{lscape}
\usepackage{makeidx}
\usepackage{mathrsfs}
\usepackage{mathtools}
\usepackage{multirow}
\usepackage{rotating}
\usepackage{textcomp}
\usepackage{xcolor}
\usepackage{url}
\usepackage{subcaption}
\usepackage{hyperref}

\DeclareGraphicsExtensions{.pdf,.png,.jpg}
\newcolumntype{L}[1]{>{\raggedright\let\newline\\\arraybackslash\hspace{0pt}}m{#1}}
\newcolumntype{C}[1]{>{\centering\let\newline\\\arraybackslash\hspace{0pt}}m{#1}}
\newcolumntype{R}[1]{>{\raggedleft\let\newline\\\arraybackslash\hspace{0pt}}m{#1}}

\newcommand{\argmax}[1]{\underset{#1}{\arg\!\max}}

\newcommand{\mPerformanceItemPileOn}{\text{IPO}}

\newcommand{\mSetStations}{\mathcal{M}}

\newcommand{\mSetOStations}{\mSetStations^{O}}

\newcommand{\mParTimeOStationHandleUnitConstant}{T^{O}}

\newcommand{\mParTimeOStationPickItemConstant}{T^{P}}

\newcommand{\mUBTimeDriveInbound}{T^{DI}}
\newcommand{\mUBTimeDriveOutbound}{T^{DO}}

\newcommand{\mUBTimeTurnOutbound}{T^{TO}}
\newcommand{\mUBTimeMoveUpNextRobot}{T^{MU}}

\definecolor{darkgrey}{RGB}{28,28,28} 
\definecolor{mediumgrey}{RGB}{71,71,71} 
\definecolor{lightgrey}{RGB}{115,115,115} 
\definecolor{lightlightgrey}{RGB}{206,206,206} 
\definecolor{darkblue}{RGB}{79,101,140} 
\definecolor{mediumblue}{RGB}{112,144,200} 
\definecolor{lightblue}{RGB}{140,180,250} 
\definecolor{darkgreen}{RGB}{41,115,46} 
\definecolor{mediumgreen}{RGB}{65,180,73} 
\definecolor{lightgreen}{RGB}{90,250,101} 
\definecolor{mediumyellow}{RGB}{247,203,56} 
\definecolor{mediumorange}{RGB}{255,138,60} 
\definecolor{mediumred}{RGB}{219,73,55} 
\definecolor{mediumviolet}{RGB}{146,90,199} 
\definecolor{mediumturquoise}{RGB}{87,189,227} 

\begin{document}


\title{Decision Rules for\\Robotic Mobile Fulfillment Systems}
\author{M. Merschformann \and T. Lamballais \and M.B.M. de Koster \and L. Suhl}

\date{\today}
\maketitle

\begin{abstract}
The Robotic Mobile Fulfillment Systems (RMFS) is a new type of robotized, parts-to-picker material handling system, designed especially for e-commerce warehouses.
Robots bring movable shelves, called pods, to workstations where inventory is put on or removed from the pods.
This paper simulates both the pick and replenishment process and studies the order assignment, pod selection and pod storage assignment problems by evaluating 
multiple decision rules per problem.
The discrete event simulation uses realistic robot movements and keeps track of every unit of inventory on every pod.
We analyze seven performance measures, e.g. throughput capacity and order due time, and find that the unit throughput is strongly correlated with the 
other performance measures.
We vary the number of robots, the number of pick stations, the number of SKUs (stock keeping units), the order size and whether returns need processing or not.
The decision rules for pick order assignment have a strong impact on the unit throughput rate. 
This is not the case for replenishment order assignment, pod selection and pod storage.
Furthermore, for warehouses with a large number of SKUs, more robots are needed for a high unit throughput rate, even if the number of pods and the dimensions of the storage area remain the same.
Lastly, processing return orders only affects the unit throughput rate for warehouse with a large number of SKUs and large pick orders. 
\end{abstract}


\section{Introduction}\label{sec:mu_intro}
The rise of e-commerce has created the need for new warehousing systems.
Traditional, manual picker-to-parts systems work best when orders are large, i.e. consist of many SKUs so that consolidation has to be organized well.
However, e-commerce orders are typically small and e-commerce warehouses are often large as they need to contain large assortments of products, which results 
in long walking distances for the pickers.
In contrast to manual picker-to-part systems, automated parts-to-picker systems eliminate the time pickers spend traveling.
Thus, they can achieve higher pick rates.\\
The Robotic Mobile Fulfillment System (RMFS) is an automated parts-to-picker system.
Robots transport movable shelves, called ``pods'', that contain the inventory, back and forth between the storage area and the workstations.
As RMFSs eliminate picker walking time, high pick rates can be expected.
Implementations suggest that pick rates can improve substantially compared to manual picker-to-parts operations, see also \cite{A527}.
The systems are mainly used by Amazon, which bought the company that invented the RMFS, Kiva Systems, and has since deployed it in its warehouses (see \cite{businesswireWebsite}).
Recently, competitors such as Swisslog, Interlink, GreyOrange, Mobile Industrial Robots and Scallog have been rolling out their versions of an RMFS.\\
The RMFS is described in more detail in \cite{A10} and \cite{A11}.
They mention that numerous operational decisions problems are yet to be examined in depth, for example the assignment of customer orders to workstations or of pods to a storage locations.
Each of these decision problems comes with a trade-off.
An order may be assigned to a workstation if it is nearing its due time, but assigning another order that has lines in common with other orders assigned to 
that workstation may result in more picks per pod and hence a reduction in the number of pod trips.
Furthermore, assigning a pod to a storage location that is close to the workstation reduces travel time, but keeping the inventory sorted by assigning pods to favorable storage location if they are likely to be needed in the near future may reduce travel times more.\\
These trade-offs are linked to the number of robots in the system.
As an example, with more robots, more trips can be done and hence the order due times can become a more important criterion than the number of picks per pod 
when selecting a pod to be transported to a workstation.
The trade-offs are also linked to the resources and conditions in the warehouse.
For example, the more SKUs a warehouse contains, the more difficult it becomes to assign orders to pick stations in such a way that multiple products can be picked from a single pod.\\
As these examples indicate, a need exists for finding methods to address the decision problems in an RMFS, for research on the performance of RMFSs across 
performance measures, and for examining performance while varying aspects like the number of robots.
This paper addresses this need.
We study the pick order assignment, replenishment order assignment, pick pod selection, replenishment pod selection, and pod storage assignment decision 
problems and propose several decision rules for each.
To see which trade-offs in performance may exist, we use different performance measures.
Furthermore, we vary three aspects of the RMFS, namely whether or not return orders need to be processed, the size of the orders, and the number of SKUs in the 
warehouse.
This study focuses on both the pick process and the replenishment process, because a more efficient replenishment process frees up robots for pick tasks.
Lastly, the number of pick stations and the number of robots per pick station is varied.
Varying these numbers shows how many pick stations and robots are needed to provide pickers with a near continuous supply of pods.\\
Section~\ref{sec:mu_rmfs} describes the RMFS in more detail, Section~\ref{sec:mu_relatedWork} points out related work, Section~\ref{sec:mu_decisionProblems} the decision problems, Section~\ref{sec:mu_rules} the decision rules, and Section~\ref{sec:mu_simulation} describes the realistic simulation built for evaluating the decision rules, while Section~\ref{sec:mu_framework} explains the evaluation framework, Section~\ref{sec:mu_results} shows the results of the analysis, and Section~\ref{sec:mu_conclusion} provides conclusions and directions for future research.

\section{The Robotic Mobile Fulfillment System}\label{sec:mu_rmfs}
\begin{figure}[!htb]
	\centering
	\begin{subfigure}{.25\linewidth}
		\includegraphics[width = \textwidth]{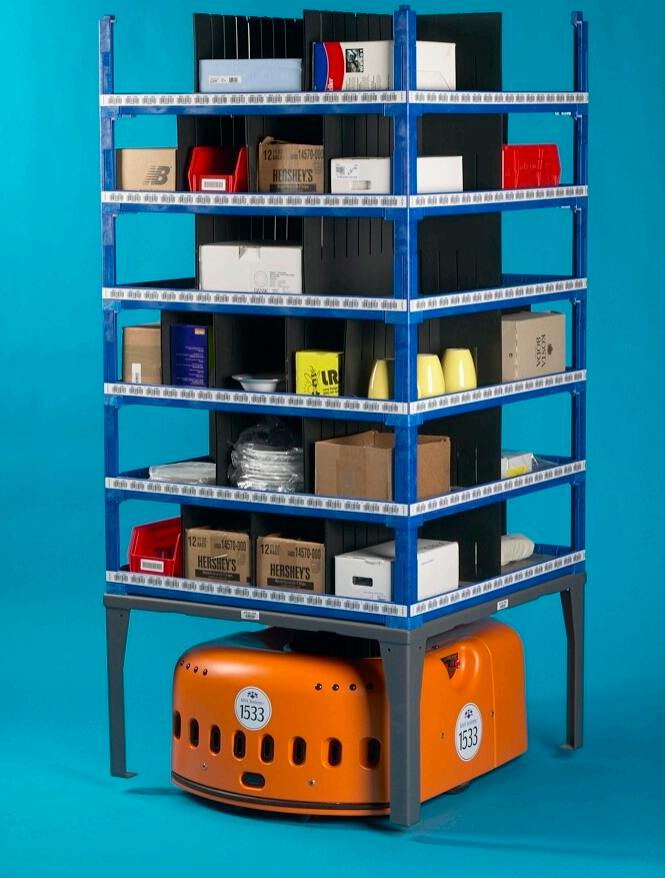}
		\caption{Robot carrying a pod (see \cite{A10})}
		\label{fig:mu_robotWithPod}
	\end{subfigure}
	\hfill
	\begin{subfigure}{.7\linewidth}
		\includegraphics[width = \textwidth]{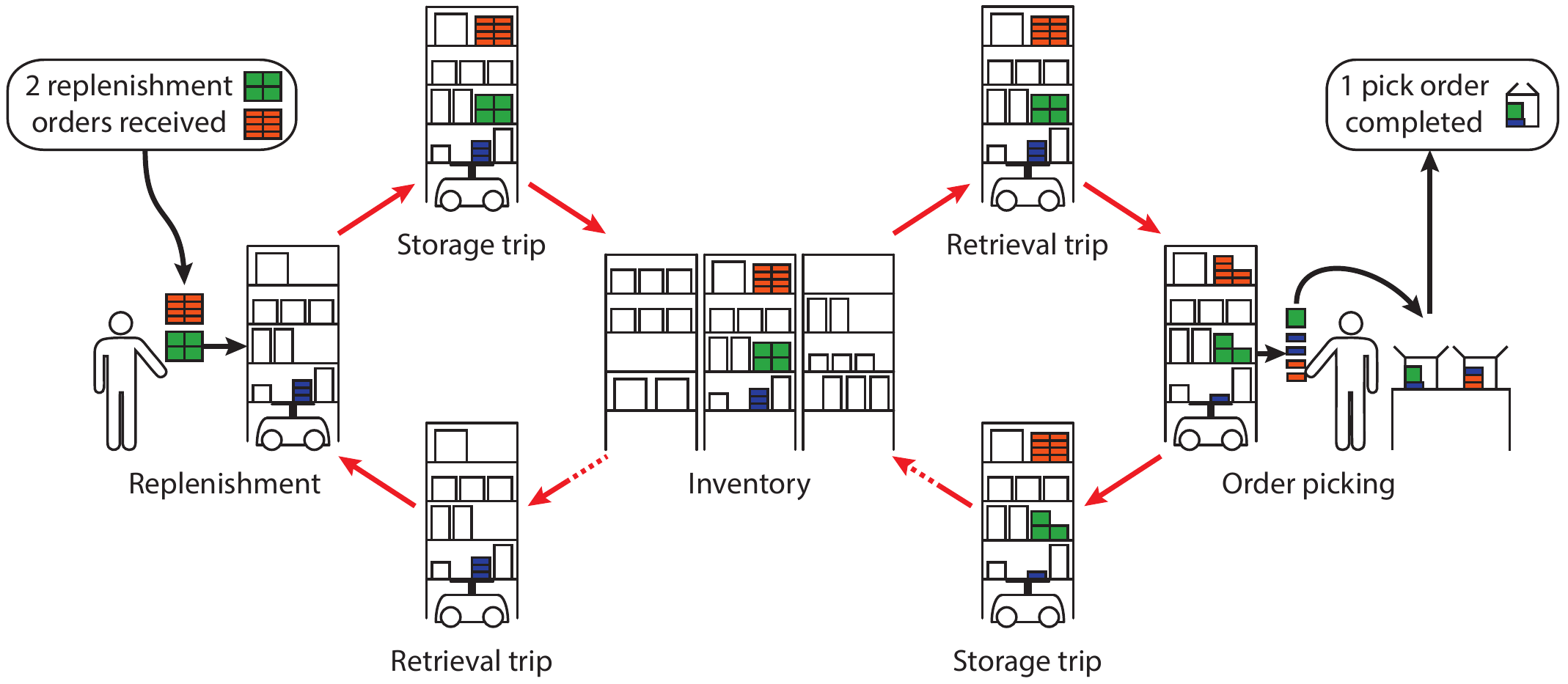}
		\caption{The internal storage / retrieval process in RMFSs (red: robot \& pod movement}
		\label{fig:mu_basicstorageretrievalprocess}
	\end{subfigure}
	\caption{The essential elements of an RMFS}
	\label{fig:mu_rmfs}
\end{figure}

An RMFS consists of shelves on which products are stored (called pods), robots that can move underneath and also carry them (see Figure~\ref{fig:mu_robotWithPod}), and work stations.
After handling a pod at a station it can be returned to a different storage location than where it was retrieved from, hence, inventory can be sorted continuously throughout the day.

\begin{figure}[!htb]
	\centering
	\includegraphics[width=0.9\textwidth]{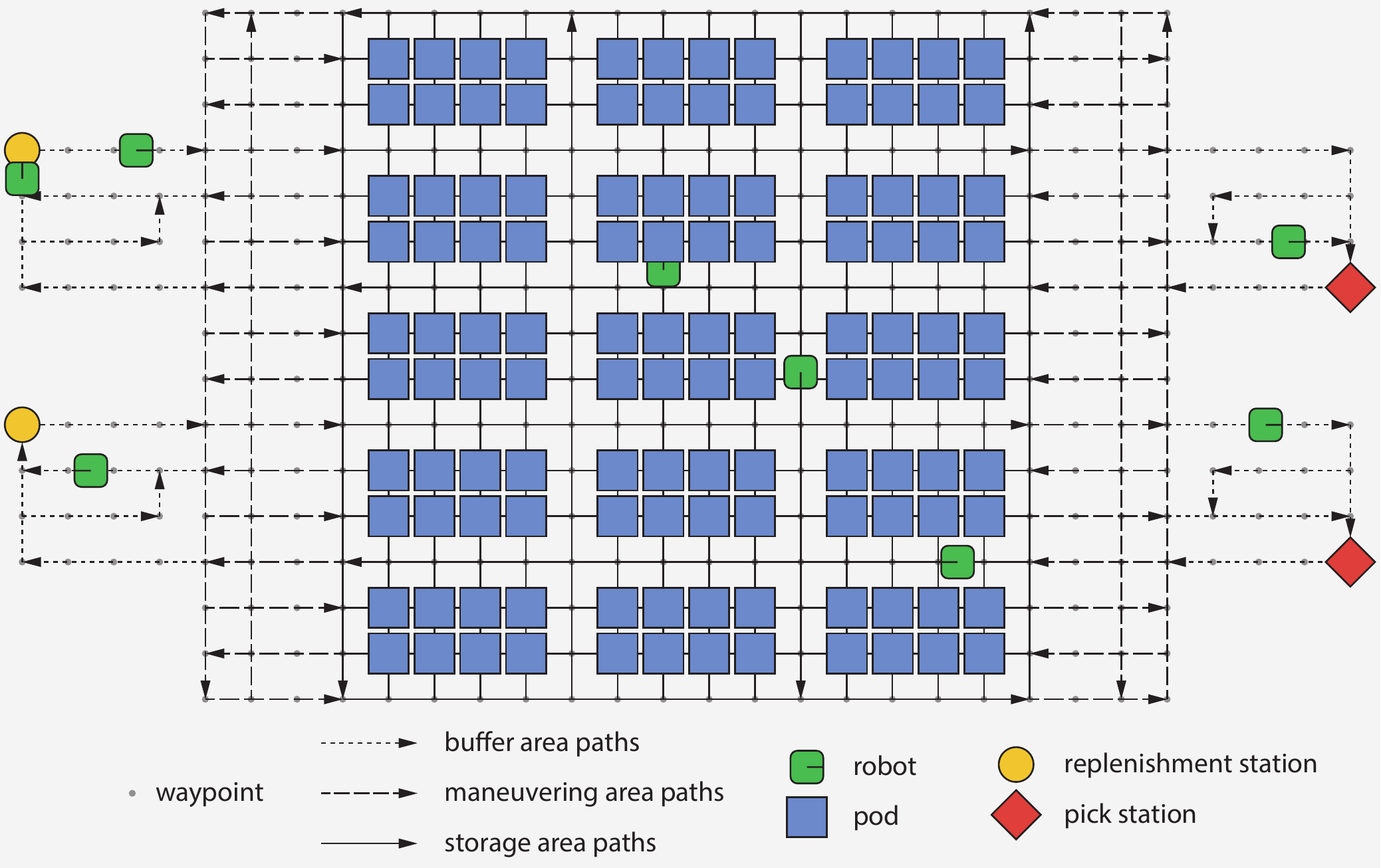}
	\caption{A top view of an RMFS layout}
	\label{fig:mu_basic-layout}
\end{figure}

Figure~\ref{fig:mu_basicstorageretrievalprocess} shows the storage and retrieval processes, where the robots transport pods between the workstations and the storage area.
Starting at the replenishment station, in the example, two replenishment orders with 4 and 8 units of two SKUs (green \& orange) are stored on a pod that was retrieved from the inventory by a robot.
The blue SKU also relevant to the process is already available on the pod in focus.
After the pod was handled at the station it is stored in inventory again.
Next, if the pod is selected for picking at a pick station, it is brought to that station.
The operator at the station then picks the units matching the open order lines at the station from the pod and puts them into the bins for the respective pick orders.
As soon as a pick order is completed it leaves the pick station and is handled by further warehouse systems.
If zoning is in place at the warehouse, the pick order may only be a part of a larger customer order and must be consolidated further with the other partial pick orders in a following sortation process.
If the customer order is already completely fulfilled at the pick station, it may be packed into a carton and prepared for shipping immediately with no further handling.
The latter may only be possible in e-commerce operations where lines per order are small.\\
Each pair of storage and retrieval trip is one robot cycle in an RMFS. During one cycle the robot does not set-down or leave the rack until it is returned to a storage location.
Note that, the pod may be brought to further replenishment or pick stations between the retrieval and the storage trip, if further replenishment or immediate picking can be done with it.
For the sake of clarity we limited the visits per cycle to one station in the example above.
While the operation of the robot is cyclic the flow of the inventory units through the system starts at a replenishment station (by storing a replenishment order) and exits at a pick station (by fulfilling a pick order). However, in contrast to other systems there is quite some overhead inventory movement, because all contained units, not only needed ones, are moved when a pod is brought to a station. The same happens during replenishment operations, if non-empty pods are moved to a replenishment station.\\
Robots navigate their paths through the warehouse using a waypoint system, which is laid out as a grid.
A path is a sequence of connected waypoints and all robots have to be guided concurrently along their paths while avoiding collisions and deadlocks.
Robots that are not carrying a pod can move underneath stationary pods and hence take other paths than robots that do carry pods, because the latter cannot use occupied storage locations.
The system layout is depicted in Figure~\ref{fig:mu_basic-layout} and consists of a storage area where the pods are stored, pick and replenishment stations 
grouped around the storage area, maneuvering areas between the storage area and the workstations, and per workstation a buffer area.
A robot carries a pod from the storage area, via the maneuvering area, to the buffer area of the destination workstation.
Only one pod is picked or replenished simultaneously.
Workers at the replenishment stations replenish the pods with new inventory.
In contrast, workers at the pick stations pick product units to fulfill orders.
A picker picks for multiple unfinished/incomplete pick orders at the same time.
For both operations the robots need to stop with a pod at a waypoint representing the access point of the respective station.
In the buffer area next to each workstation, robots carrying pods can wait for their turn.
In the middle of the layout a number of waypoints is used as possible storage locations where pods can be put when they are not used.
Every storage location is directly reachable from an aisle and access to a storage location cannot be blocked by stored pods.
Travel in the aisles is single-directional to avoid gridlock and reduce congestion.\\
The system has the ability to adapt to changing demand conditions. 
E.g., if order arrival rates of some SKUs drop, pods containing those SKUs can be relocated further away from the pick stations.
This relocation frees up storage locations near the pick stations for pods containing SKUs with high order arrival rates.
Pods can be relocated when returning from a workstation, hence the inventory can be continually sorted in response to changing demand.

\section{Related Work}\label{sec:mu_relatedWork}
To this date no detailed discrete event simulation based research has been done for RMFS. Moreover, most research on RMFSs to date uses queueing networks to study design questions on the strategic level. This work aims to close the gap by delivering insights about RMFS using a very detailed simulation framework that integrates most dynamic effects an operator faces. Next, we first outline the queuing network based research and close this section with simulation based work.\\
\cite{A778} create queueing networks similar to earlier queueing networks used for autonomous vehicle storage and retrieval systems (AVS/RS) and automated storage and retrieval systems (AS/RS) (see \cite{A442} and \cite{A70}).
Their queueing networks capture both pick and replenishment operations but cannot model robot movement realistically.
They estimate the order throughput time for single-line orders.
\cite{A677} create a different queueing network for both single- and multi-line orders, with and without zoning in the storage area, that captures only the pick operations, but that does include realistic robot movement.
Their model can accurately estimate the expected order cycle time, workstation utilization and robot utilization.
\cite{A677} determine how the storage area dimensions and the workstation placement around the storage area affect the maximum order throughput, by evaluating 
a large number of possible designs.
\cite{secondPaper} develop a queueing network that addresses problems on a tactical level.
They show the effect of the number of pods per SKU and of the replenishment level of a pod on order throughput, and they show what the optimal ratio of the 
number of pick stations to the number of replenishment stations is.
They find that it is better to replenish pods before they are entirely empty, even with multiple pods per SKU.
\cite{Zou.2017} use semi-open queueing networks to analyze the policy for assigning robots to pick stations.
The authors find that the random policy is significantly outperformed by the proposed handling-speeds-based assignment rule when facing varying service rates of the pickers.
\cite{Zou.2017b} build a semi-open queueing network for evaluating the effects of battery management in RMFS.
The strategies of battery swapping, automated plug-in charging and inductive charging at the pick station are compared.
The authors come to the conclusion that battery swapping is generally more expensive than plug-in charging while inductive charging outperforms both in throughput and costs, if robot prices and retrieval times are low.\\
\cite{A10} and \cite{A11} mention several decision problems on the operational level that they encountered in practice.
One of the few studies that address decision problems on the operational level is by \cite{A810}.
They provide methods for optimally batching the pick orders and sequencing both the pick orders and the pods transported to the stations. 
They show that an optimized pick order processing requires only half the number of robots that a pick order process based on simple decision rules would need.
\cite{Roodbergen.2014} utilize a simulation based approach in order to optimize the warehouse layout of a manual order picking system for an industrial partner.
The authors devise an integrated approach taking on to certain design decisions as well as selecting control policies.
The simulation is thereby used ``as a solution tool and an evaluation system'' (see \cite{Roodbergen.2014}).
\cite{A618} use a simulation based approach for evaluating the performance of policy sets for manual order picking systems.
The authors make use of DEA as a tool for obtaining a comparable performance indicator among the policy sets.
\cite{Beckschafer.2017} use a discrete event simulation approach similar to this work for assessing storage policies for Automated Grid-based Storage systems.
The authors find that even simple strategies improve the system efficiency, which encourages research on more complex strategies.
\cite{projectDelta} develop a Markow decision process (MDP) model for addressing the resource reallocation problem, i.e., the problem of deciding how many workers and robots to allocate to the pick process and replenishment process continually throughout time.
The assumptions related to replenishment differ strongly across the papers mentioned above, and the number of approaches to replenishment in practical applications is diverse as well.

\section{Decision Problems}\label{sec:mu_decisionProblems}
This section introduces the decision problems considered in this paper and places them within the context of other decision problems in an RMFS.
Requests to the system occur via pick orders or replenishment orders.
Upon receipt, pallets are broken up into smaller parts consisting of multiple units of one SKU.
A replenishment order is a request to place one such part, i.e. a number of units of one specific product, on a pod.

\begin{figure}[!tb]
	\centering
	\includegraphics[width = 0.8\textwidth]{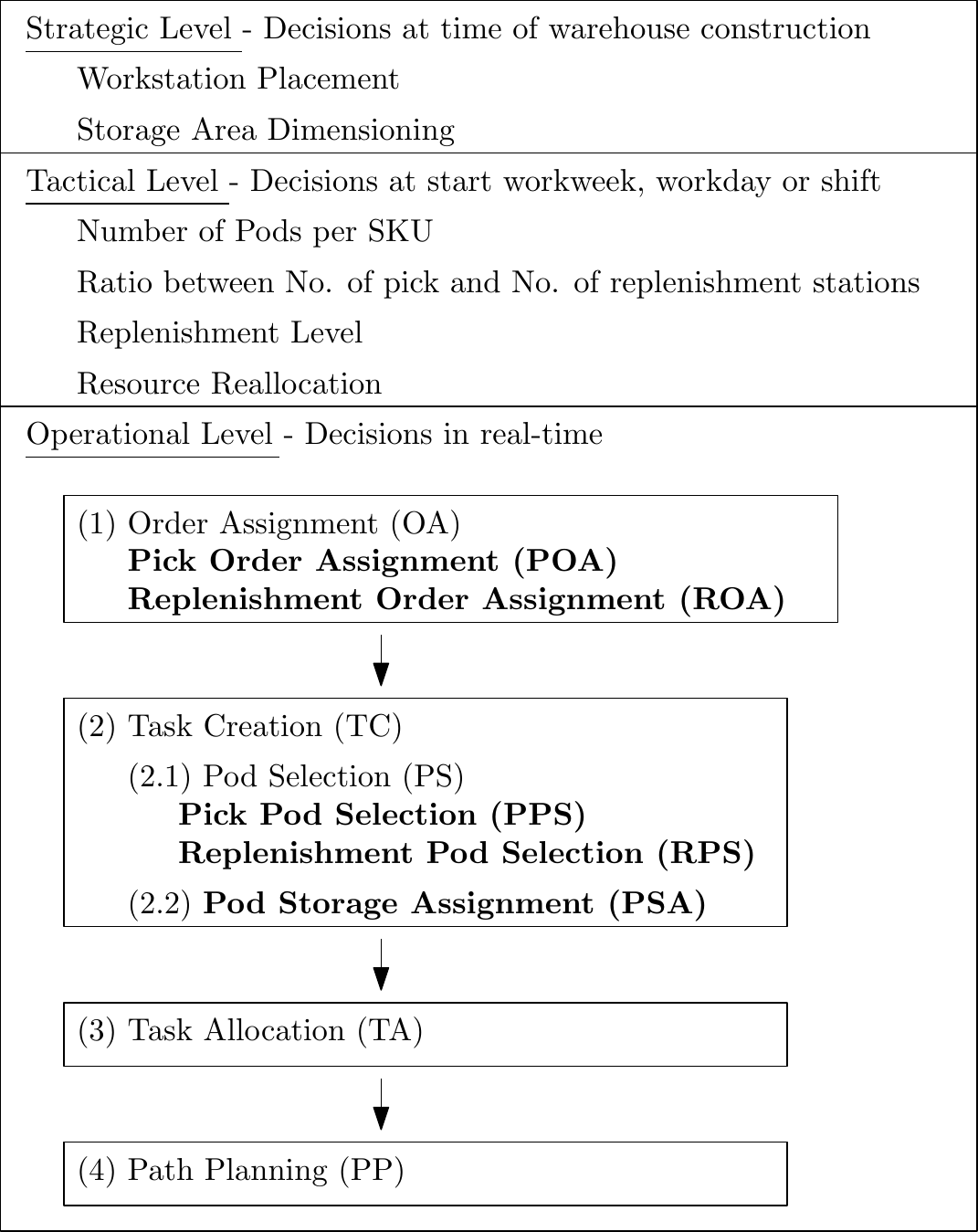}
	\caption{Hierarchical overview of the decision problems and their relations}
	\label{fig:mu_decisionProblemHierarchy}
\end{figure}

We structure the decisions at the operational level in four steps: 
(1) Order Assignment (OA), the assignment of pick or replenishment orders to workstations, 
(2) Task Creation (TC), the creation of tasks for the robots, 
(3) Task Allocation (TA), the allocation of tasks to robots, and 
(4) Path Planning (PP), the creation of paths along which the robots will move.
There are two kinds of Order Assignment decisions: the assignment of pick orders to pick stations, called the Pick Order Assignment (POA) problem, and the 
assignment of replenishment orders to replenishment stations, called the Replenishment Order Assignment (ROA) problem.
In the second step, a task is defined as transporting a specific pod to a specific workstation and back to a specific storage location.
Therefore, for each workstation, the Task Creation decision problem includes the two subproblems of 
(2.1) deciding which pod to select for transportation, the Pod Selection (PS) decision problem, and 
(2.2) deciding at which storage location to return the pod, the Pod Storage Assignment (PSA) decision problem.
The Pod Selection (PS) decision problem differs for the pick and replenishment process, because for the pick process the due times of the pick orders is important in selecting a pod.
Pod selection in the pick process is called Pick Pod Selection (PPS) and pod selection in the replenishment process is called Replenishment Pod Selection (RPS).
Task Creation uses the pick order and replenishment order assignments to select suitable pods and subsequently converts the requests for the selected pods into tasks for pod transportation between the workstations and the storage area.
Task Allocation creates a trip by building a sequence of tasks for the robots to execute.
These sequenced tasks implicitly define trips and serve as input for the Path Planning algorithms, where a path is generated for a robot to follow.\\
Figure~\ref{fig:mu_decisionProblemHierarchy} shows an overview of the decision problems at the strategic, tactical and operational level in an RMFS, with the 
problems addressed in this paper in bold.
As can be seen in Figure~\ref{fig:mu_decisionProblemHierarchy}, this paper focuses on decision problems at the operational level.
We use the term ``decision rule'' to refer to a fairly simple method to solve a decision problem.
The aim of this paper is to evaluate several decision rules per decision problem.
Some decision rules may closely resemble common best practices, whereas others may be more specific to RMFS.
The Task Allocation decision problem is intertwined with the Path Planning decision problem, which has been addressed by \cite{A708}.
Therefore we do not consider the Task Allocation and Path Planning decision problems.
We do address Pick Order Assignment (POA), Replenishment Order Assignment (ROA), Pick Pod Selection (PPS), Replenishment Pod Selection (RPS), and Pod Storage Assignment (PSA).
For Pick Order Assignment, we assume there is a constant backlog, and the pick stations are always filled to full capacity with pick orders.
Whenever a pick order is fulfilled and leaves its pick station, a pick order has to be selected from the backlog and assigned to the pick station.
For replenishment orders, we assume that the sequence of replenishment orders inbound to the system cannot be altered anymore.
This assumption resembles the situation in conventional conveyor-based material handling components that do not allow sequence modification but only load routing.
Moreover, we aim to avoid taking decision problems outside of the system's boundaries into account, e.g., different dispatching rules of preceding systems.
The replenishment stations have a finite capacity.
If a replenishment order arrives and multiple replenishment station have capacity left, the ROA decision rule determines to which replenishment station the replenishment order is assigned.
If no place is available, replenishment orders are put in a replenishment order backlog.
When a replenishment order is fulfilled at one of the replenishment stations, a new replenishment order is chosen from the replenishment order backlog according to the FCFS rule.
Table~\ref{tab:mu_decisionProblems} summarizes the decision problems addressed in this paper.\\
At this point we also introduce the concept of ``pile-on'' (sometimes also called ``hit-rate'').
Pile-on as a concept refers to the average number of units that are picked from a pod every time a pod is presented to a picker at a pick station.
Pile-on as a metric measures the number of units (across all SKUs) picked from a pod when presented to a picker at a pick station, averaged across every visit 
of a pod to a pick station during the entire time horizon.
In other words, pile-on is measured in ``units picked per pod visit to a pick station''.
The higher the pile-on is, the fewer pods need to be transported between the pick stations and the storage area, which may reduce the number of robots needed.

\begin{table}[!htbp]
	\caption{Decision Problems}
	\label{tab:mu_decisionProblems}
	\begin{center}
		\resizebox{\textwidth}{!}{
		\begin{tabular}{p{1cm}p{3.5cm}p{5cm}p{7cm}}
			\hline
			Abb. & Name & Description & Trigger\\
			\hline
			POA & Pick Order \newline  {Assignment} & Choosing a pick order from the backlog & When another pick order is fulfilled and leaves the pick station, creating room for the next pick order to be assigned\\
			ROA & Replenishment \newline Order {Assignment} & Selecting the replenishment station for the next replenishment order & When a replenishment order arrives at the system and one or more replenishment stations have capacity left\\
			PPS & Pick Pod {Selection} & Selecting a pod to transport to a pick station & When a robot working for a pick station needs a new task\\
			RPS & Replenishment Pod {Selection} & Select a pod for the next replenishment order & Depends on the ROA decision rule\\
			PSA & Pod Storage\newline  {Assignment} & Choosing a storage location for a pod & When a pod leaves a workstation\\
			\hline
		\end{tabular}
		}
	\end{center}
\end{table}

\section{Decision Rules}\label{sec:mu_rules}
To solve the operational problems, we define several decision rules per decision problem that are evaluated in a realistic simulation.
Several Path Planning algorithms for the RMFS are compared in \cite{A708}, therefore this decision problem will not be addressed in this paper. Thus, we selected WHCA$^*_v$, one of the best performing algorithms from the paper, as the path planning engine for this work.
Additionally, we fix the Task Allocation algorithm to a simple method that first assigns two-thirds of the robots to pick operations and the rest to 
replenishment operations.
Then, it aims to equally distribute the robots across the respective stations.
This means a robot will only do tasks related to the station it is assigned to.
This section will therefore only describe decision rules for the Pick Order Assignment, Replenishment Order Assignment, Pick Pod Selection, Replenishment Pod Selection and Pod Storage Assignment decision problems.\\
While replenishment and pick operations are similar in the sense that high throughput should be achieved with few resources, the main asymmetry between both is 
that for the former the goal is to fill the inventory as quickly as possible and for the latter to empty it as quickly as possible.
This means that for replenishment operations we aim to replenish pods fast to have them available for pick operations early while preparing pod content such that it allows for a high pile-on during pick operations.
For pick operations we aim to achieve a high pile-on and keeping trips short to fulfill as many orders as possible while also considering due times of the pick orders.
Furthermore, we do not allow the sequence of replenishment orders to be modified.
In contrast, for pick orders we allow to arbitrarily choose one order from the backlog.
Lastly, pick orders have due times.
All of this leads to different strategies we focus on per decision problem, instead of fully symmetric rules between pick and replenishment decision problems.\\
For a more precise description of some of the rules we introduce the notation shown in Table~\ref{tab:mu_rule_definition_symbols}.

\begin{table}[htb]
	\caption{Overview of the symbols used in the rule descriptions}
	\label{tab:mu_rule_definition_symbols}
	\setlength{\tabcolsep}{2mm}
	\centering
	\begin{tabular}{l|l}
		Symbol & Explanation \\
		\hline
		$\mathcal{P}$ & Set of all pods \\
		$\mathcal{P}^I_s$ & Set of pods heading to station $s$ \\
		$\mathcal{I}$ & Set of all SKUs \\
		$\mathcal{O}^B$ & Set of pick orders in backlog \\
		$\mathcal{O}^S_s$ & Set of pick orders assigned to station $s$ \\
		$C(p,i)$ & Number of units of SKU $i$ contained in pod $p$ \\
		$L(o,i)$ & Required units necessary to fulfill line $i$ of order $o$ \\
		$D(o,i)$ & Remaining units necessary to fulfill line $i$ of order $o$ \\
		$t^D_o$ & Due time of order $o$ \\
		$t^S_o$ & Time of assignment to the station of order $o$ \\
		$t$ & Time of deciding \\
	\end{tabular}
\end{table}

\subsection{Pick Order Assignment Rules}

A pick station has to be chosen for every pick order submitted to the system and the pick order itself has to be chosen from the order backlog.
In this work, we consider a pick order backlog of constant size, i.e., as soon as an order is removed from the backlog a new one is generated to replace it.
This and the immediate replacement of orders completed at a station lead to only one option available to assign any pick order to: the slot of the just completed order.
Hence, the choice of station is not a degree of freedom in this work.
The rare occasions of multiple orders to be completed at the same time are handled by assigning the orders to the pick stations randomly.
Hence, we only investigate rules for selecting the next pick order from the backlog to fill the only open slot at a station.
We devise six rules to solve this problem: ``Random'', ``FCFS'', ``Due-Time'', ``Fast-Lane'', ``Common-Lines'' and ``Pod-Match'':

\begin{description}
	\item[Random] The Random rule randomly selects a next pick order from the backlog and is used as a benchmark.
	\item[FCFS] The FCFS rule assigns the pick order that was first received.
	The rationale behind this is to keep pick order throughput times short.
	\item[Due-Time] The Due-Time rule selects the pick order with the earliest due time from the backlog and assigns it to a station.
	This is a greedy approach aiming to finish the pick orders before their deadline.
	\item[Fast-Lane] The Fast-Lane rule randomly selects a pick order from the backlog like the Random rule, but keeps one slot at each pick station open for immediately completeable pick orders.
	I.e., only pick orders ($o$), for whom all lines and all units of inventory are available on the next pod ($p_n$) will be assigned to this station's ``fast-lane'' order slot (see Equation~\eqref{eq:rulefastlane}).
	Thus, orders assigned to the ``fast-lane'' slot are processed shortly after assignment.
	The next pod of the station is either a not completely processed pod the picker is currently working on or the next pod in the station's queue, if no such pod is available.
	In cases where no pod reached the station's queue yet, we consider the pod with the shortest remaining path to estimate the next pod.
	When facing multiple options we use a random tie-breaker.
	Note that this rule can be combined with any other proposed POA rule.
	The reason we combine it with random selection is to better assess the impact of the idea itself.
	\begin{equation}\label{eq:rulefastlane}
		\forall i \in \mathcal{I}: L(o,i) \le C(p_n,i)
	\end{equation}
	\item[Common-Lines] The Common-Lines rule compares the station's ($s$) currently assigned pick orders with all orders from the backlog and selects the one with most lines in common for assignment (see Equation~\eqref{eq:rulecommonlines}).
	The rationale behind this is to increase pile-on by exploiting synergies among the pick orders.
	When facing multiple options we use a random tie-breaker.
	\begin{equation}\label{eq:rulecommonlines}
		\argmax{o \in \mathcal{O}^B} \sum_{o' \in \mathcal{O}^S_s} \sum_{i \in \mathcal{I}} \left( \begin{cases}
		1 & L(o,i) > 0 \wedge L(o',i) > 0 \\
		0 & otherwise
		\end{cases} \right)
	\end{equation}
	\item[Pod-Match] The Pod-Match rule selects the pick order from the backlog that matches best the pods heading to the station ($s$) at the moment of assignment 
	best. 
	I.e., the more units of the pick order are already available in the pods the better the match (see Equation~\eqref{eq:rulepodmatch}).
	When facing multiple options we use a random tie-breaker.
	\begin{equation}\label{eq:rulepodmatch}
		\argmax{o \in \mathcal{O}^B} \sum_{p \in \mathcal{P}^I_s} \sum_{i \in \mathcal{I}} \left( \min \left( C(p,i), D(o,i) \right) \right)
	\end{equation}
\end{description}

\subsection{Replenishment Order Assignment Rules}

As a result of the assumptions that replenishment orders arrive in a fixed sequence, we investigate only two different approaches for assigning replenishment orders to the stations, i.e., immediate Random assignment and batching of customer orders that go on the same pod.
Hence, we construct two rules for replenishment assignment: ``Random'' and ``Pod-Batch'':

\begin{description}
	\item[Random] The Random rule randomly selects a next station with sufficient remaining capacity to allocate incoming replenishment orders to. If no such station is available, the order will wait until one becomes available again.
	\item[Pod-Batch] The Pod-Batch rule tries to use a pod already selected to go to a replenishment station for assigning the next replenishment order.
	In other words, the Pod-Batch rule first waits for the Replenishment Pod Selection (Section~\ref{sec:rps}) rule to decide which orders are assigned to which pod, and then uses the same replenishment station for the orders of one pod. 
	If the replenishment orders do not fit one station, they wait until a station with sufficient capacity becomes available.
	During this time all consecutive orders are also blocked, because the sequence cannot be altered.
\end{description}

\subsection{Pick Pod Selection Rules}

Every time a robot working for a pick station $s$ requests a next task, a pod suitable for picking at pick station $s$ must be selected.
We require for all rules that at least one unit can be picked from the pod.
This means that no pod is brought to a station completely in vain and additionally it implies a pile-on of at least 1.
The six PPS rules used in this paper are the ``Random'', ``Nearest'',  ``Pile-on'', ``Demand'', ``Lateness'', and ``Age'' rules:

\begin{description}
	\item[Random] The Random rule randomly selects a pod that offers at least one useful unit for picking.
	\item[Nearest] The Nearest rule selects the pod which has the least estimated path time towards the station according to the path planning algorithm and 
	that offers at least one useful unit for picking.
	\item[Pile-on] The Pile-on rule selects the pod that offers most units necessary to fulfill the orders at the station (see Equation~\eqref{eq:rulepileon}). Ties are broken by favoring pods with which more orders can be completed. If ties still persist, they are broken randomly.
	\begin{equation}\label{eq:rulepileon}
		\argmax{p \in \mathcal{P}} \sum_{i \in \mathcal{I}} \sum_{o \in \mathcal{O}^S_s} \left( \min \left( C(p,i), D(o,i) \right) \right)
	\end{equation}
	\item[Demand] The Demand rule selects the pod whose content is most demanded considering the current pick order backlog situation, i.e. the pod with most 
	units demanded in the backlog is chosen (see Equation~\eqref{eq:ruledemand}). Ties are broken randomly.
	\begin{equation}\label{eq:ruledemand}
		\argmax{p \in \mathcal{P}} \sum_{i \in \mathcal{I}} \sum_{o \in \mathcal{O}^B} \min \left( C(p,i), D(o,i) \right)
	\end{equation}
	\item[Lateness] The Lateness rule aims to finish late pick orders by selecting a pod that offers units needed to fulfill open order lines with most lateness at the station, i.e., for one order the time the order is late is summed as fractions of the open picks (see Equation~\eqref{eq:rulelateness}). If no order is late, the resulting ties are broken by using the same metric but replacing $\max \left( t - t^D_o , 0 \right)$ with $t^D_o$, thus, selecting pods for orders whose due times are most imminent.
	\begin{equation}\label{eq:rulelateness}
		\argmax{p \in \mathcal{P}} \sum_{i \in \mathcal{I}} \sum_{o \in \mathcal{O}^S_s} \left( \frac{\min \left( C(p,i), D(o,i) \right) }{\sum_{i' \in \mathcal{I}} D(o,i') } \max \left( t - t^D_o , 0 \right) \right) 
	\end{equation}
	\item[Age] The Age rule aims to finish the oldest pick orders of a station by selecting a pod that offers units needed to fulfill the oldest open order lines, i.e. for one order the time the order spent assigned to the station is summed as fractions of the open picks (see Equation~\eqref{eq:ruleage})
	\begin{equation}\label{eq:ruleage}
		\argmax{p \in \mathcal{P}} \sum_{i \in \mathcal{I}} \sum_{o \in \mathcal{O}^S_s} \left( \frac{\min \left( C(p,i), D(o,i) \right) }{\sum_{i' \in \mathcal{I}} D(o,i') } \left( t - t^S_o \right) \right) 
	\end{equation}
\end{description}

\subsection{Replenishment Pod Selection Rules}\label{sec:rps}

For every replenishment order, a suitable pod with sufficient remaining storage capacity needs to be chosen.
The decision is taken right before the replenishment order is assigned to a replenishment station.
Depending on the selected ROA and RPS rules both are either invoked simultaneously or, if there is a dependency between the two, one after the other.
An example for the latter case is the combination of the PodBatch ROA rule with the Emptiest RPS rule, because the PodBatch rule relies on an already selected pod for the replenishment order.
Since Replenishment Pod Selection determines the composition of the pods, it offers many possibilities to create pods with different features, e.g. high frequency pods that combine frequently ordered products, or family-based pods combining products that are often ordered together. 
If all replenishment orders assigned to the same pod are assigned to the same replenishment station, only one trip is necessary to place all replenishment orders on the pod, which reduces the number of robot movements.\\
The five RPS rules used in this paper are the ``Random'', ``Emptiest'', ``Nearest'', ``Least-Demand'' and ``Class'' rules:

\begin{description}
	\item[Random] The Random rule selects a random pod with sufficient remaining capacity.
	\item[Emptiest] The Emptiest rule assigns replenishment orders to the 
	emptiest pod and reuses the same pod for subsequent replenishment orders until it is full or used at a station.
	\item[Nearest] The Nearest rule assigns an incoming replenishment order to the nearest pod with sufficient remaining capacity.
	\item[Least-Demand] With the Least-Demand rule an incoming replenishment order is assigned to the pod currently offering the least demanded inventory, i.e. the pod with the least units offered when compared to the aggregated demand by assigned and backlogged pick orders is selected. Thus, this pod is not useful for pick-operations at the time of selection and by this it is not disadvantageous to block it for replenishment operations.
	\item[Class] The Class rule assigns incoming replenishment orders to a pod of the same class as the replenishment order, i.e. fast moving SKUs to pods with other fast moving SKUs. 
	The classes are built by a background mechanism for which the cumulative relative amount of pods per class are given. 
	For this work we use ``$0.1$, $0.3$, $1.0$'', i.e., three classes where the first class holds 10 \% of the pods for the highest frequency SKUs, the second class holds 20 \% and the last class holds the remaining ones, which are the ones with the lowest frequency SKUs.
	To assign a replenishment order of a certain class, the emptiest pod is selected from the pods of that particular class.
	Similar to the Emptiest rule, a selected pod is used for the subsequent incoming replenishment orders of the same class until no more replenishment orders fit the pod or until the respective pod completes its visit to a replenishment station.
\end{description}

\subsection{Pod Storage Assignment Rules}

For each pod an unoccupied storage location has to be selected, every time after visiting a pick or replenishment station. 
PSA is an important aspect of the RMFS, because being able to change the storage location of pods after every visit to a workstation is what makes continuous automatic sorting possible.
For PSA, five decision rules are examined, namely the ``Random'', ``Fixed'', ``Nearest'', ``Station-Based'' and ``Class'' rules.

\begin{description}
	\item[Random] The Random rule chooses a random free storage location.
	\item[Fixed] The Fixed rule maintains the initially assigned storage location for all pods.
	\item[Nearest] The Nearest rule stores pods at the nearest unoccupied storage location in terms of shortest estimated path time. This path time is 
	determined using an A$^*$ algorithm that takes the time needed for turning the robot (with or without pod) into account.
	\item[Station-Based] The Station-based rule is a variant on the Nearest rule, i.e. instead of bringing the pod to a storage location that is nearest to the robot's position the storage location with shortest path time to a pick station is selected. 
	The greatest difference with the Nearest rule is in the storage locations chosen for pods returning from a visit to a replenishment station.
	\item[Class] The Class rule brings pods back to storage locations of the same class, where classes are constructed in a similar fashion as in the RPS 
	decision problem, but based on the shortest path time to a pick station. Within a class, a storage location for a pod is selected analogously to the Nearest rule.
\end{description}

Table~\ref{tab:mu_rules} provides an overview of the decision rules per decision problem and shows how the decision rules are labeled across decision problems.
Note that choosing a rule for one decision problem may jeopardize strategies chosen for others. 
For example, a random Pick Order Assignment may have a negative impact on a Turnover-based approach for assigning replenishment orders to storage locations, because it does not respect the units currently positioned near the pick station while assigning orders to it. 
Hence, a selection respecting mutual influences has to be done to provide an efficient compilation of rules that is able to adequately overcome the planning problems in such a system.

\begin{table}[htb]
	\caption{Overview of the Decision Rules per Decision Problem}
	\label{tab:mu_rules}
	\centering
	\setlength{\tabcolsep}{1mm}
	\resizebox{\textwidth}{!}{
	\begin{tabular}{r|l}
		\hline
		Decision Problem & Decision Rules\\
		\hline
		POA & Random, FCFS, Due-Time, Fast-Lane, Common-Lines, Pod-Match \\
		ROA & Random, Pod-Batch \\
		PPS & Random, Nearest, Pile-on, Demand, Lateness, Age \\
		RPS & Random, Emptiest, Nearest, Least-Demand, Class \\
		PSA & Random, Fixed, Nearest, Station-Based, Class \\
		\hline
	\end{tabular}
	}
\end{table}

\section{Simulation Framework}\label{sec:mu_simulation}
\begin{figure}[htb]
	\centering
	\begin{subfigure}[b]{0.62\textwidth}
		\includegraphics[width = \textwidth]{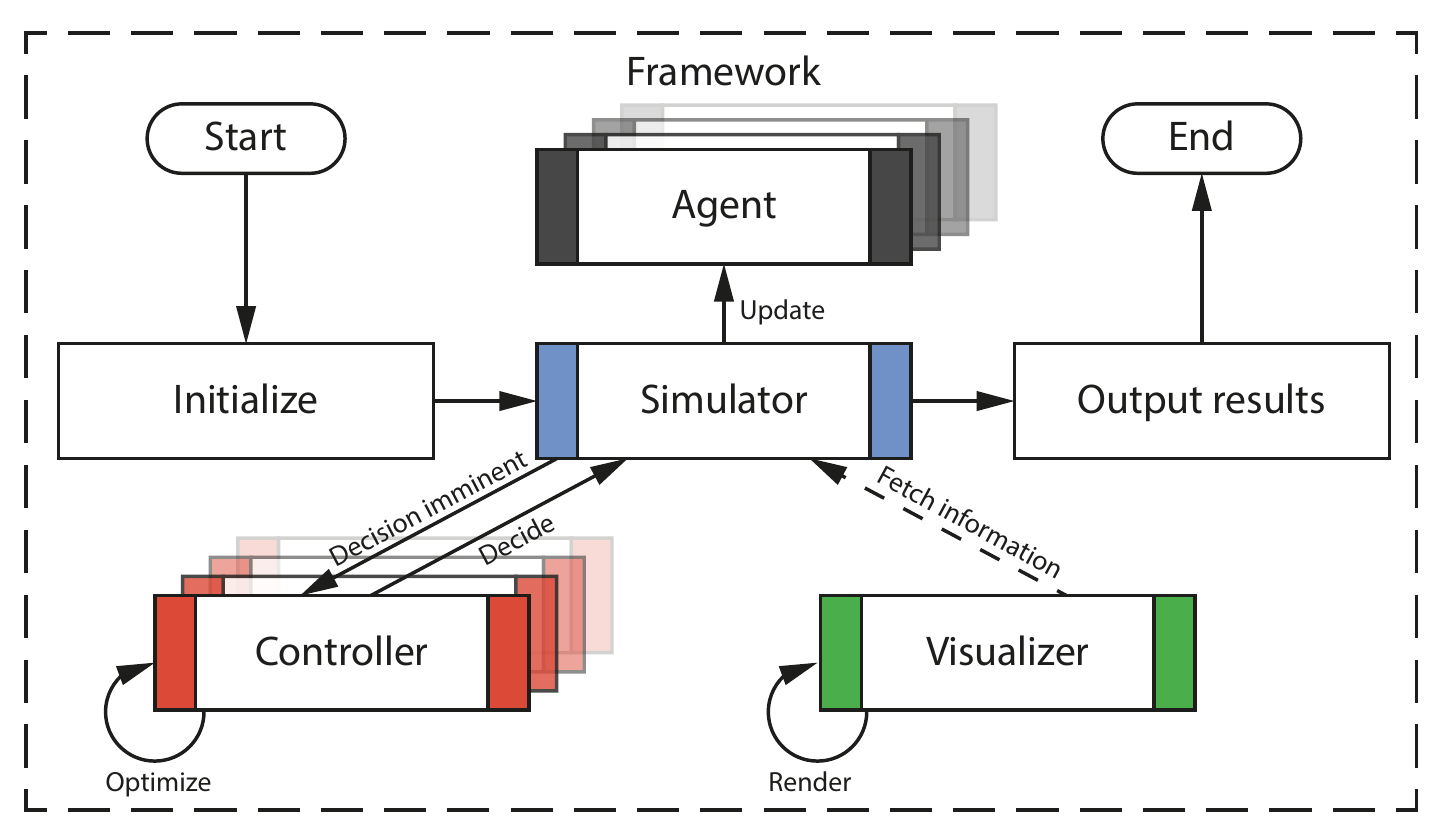}
		\caption{Overview of the simulation process.}
		\label{fig:mu_simulationframework}
	\end{subfigure}
	~
	\begin{subfigure}[b]{0.35\textwidth}
		\includegraphics[width = \textwidth]{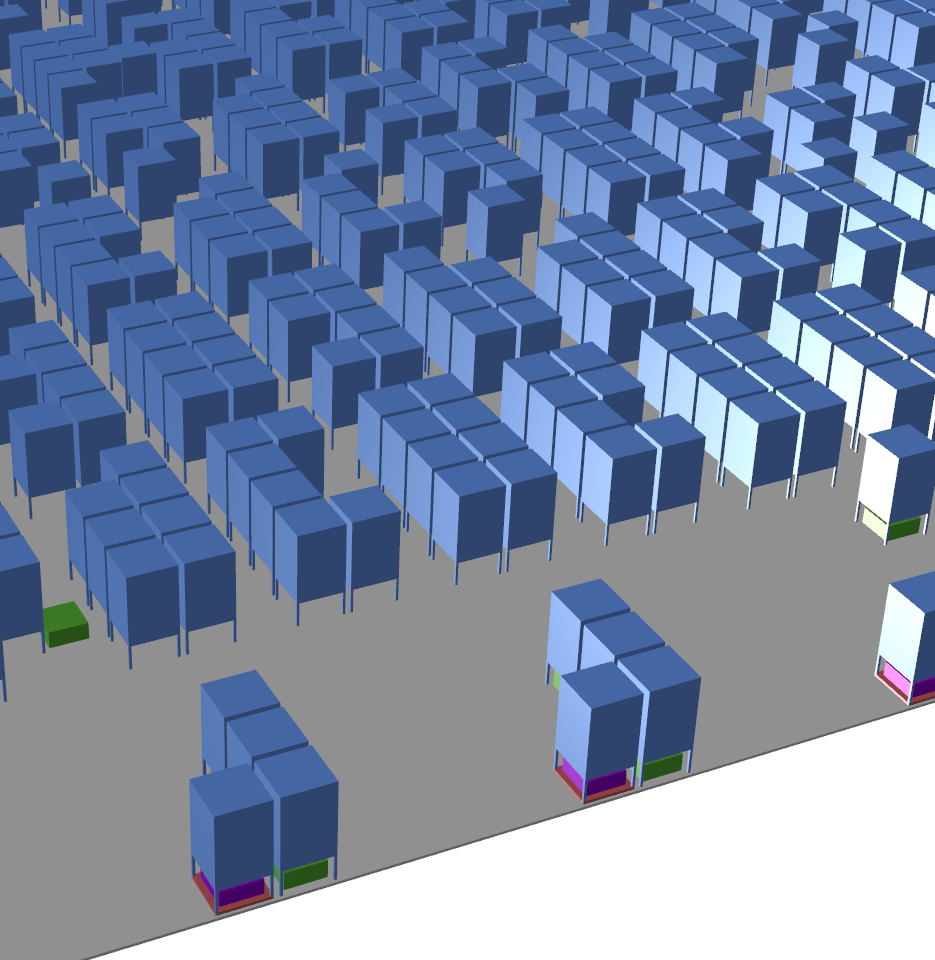}
		\caption{Visualization screenshot}
		\label{fig:mu_simulationscreenshot}
	\end{subfigure}
	\caption{RAWSim-O simulation framework}
\end{figure}

\begin{figure}[htb]
	\centering
	\includegraphics[width = 0.85\textwidth]{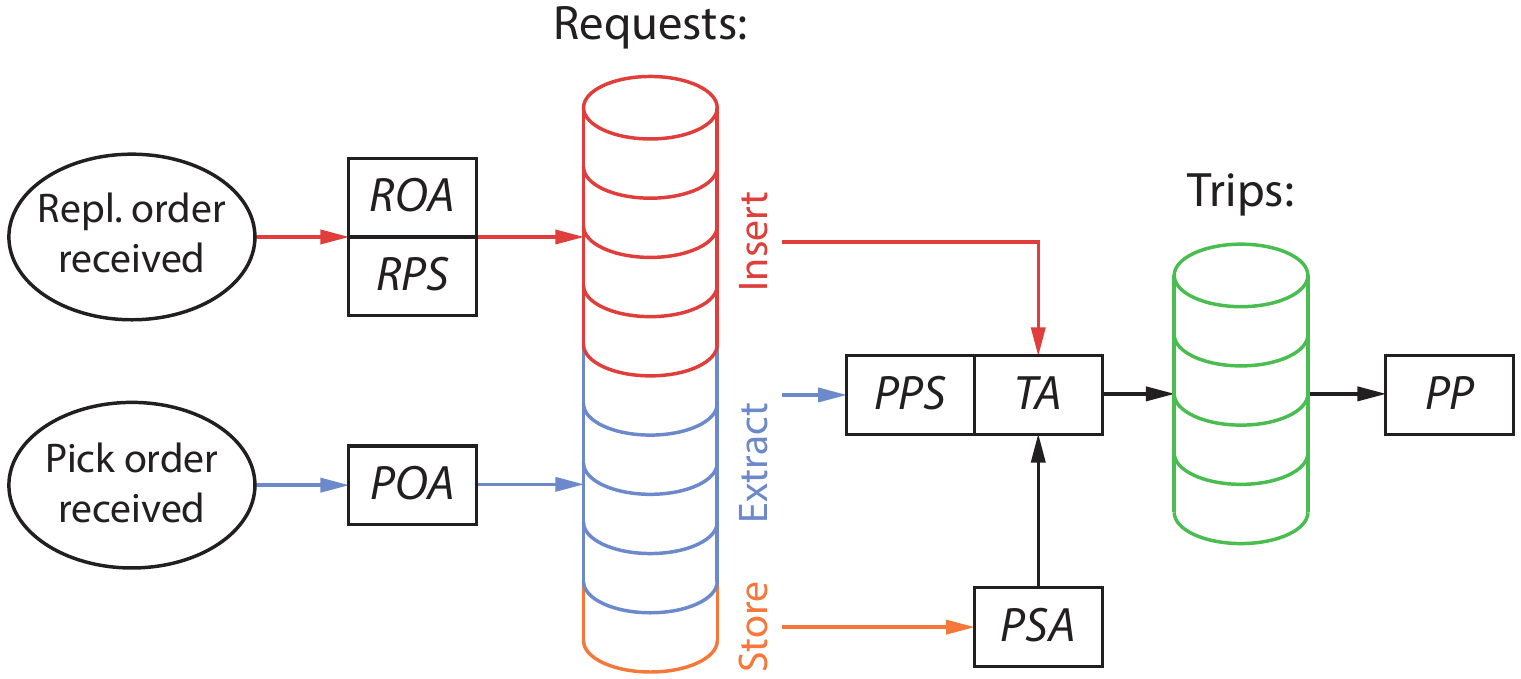}
	\caption{Order of decisions to be done induced by receiving pick and replenishment orders}
	\label{fig:mu_problemdependencies}
\end{figure}

In this work we use our simulation framework, called ``RAWSim-O'', which is inspired by the work of \cite{Hazard.2006}.
A more detailed description of the framework can be found in \cite{A813} while the source code is available at \url{https://github.com/merschformann/RAWSim-O}.
Similar to \cite{Hazard.2006}, we use an agent-based and event-driven simulation focusing at a detailed view of the system.
The basic simulation process is managed by the core simulator instance (see Figure~\ref{fig:mu_simulationframework}), which is responsible for obtaining the next event and updating the agents.
Agents can either represent real entities like robots and stations or virtual entities like process managers, e.g. for emulating order processes.
Every decision that has to be made is passed to the corresponding controller.
The controller can either immediately decide or can buffer multiple requests in order to optimize and release the decision later on.
However, in this work we only consider ad-hoc decision rules with the former approach.
To allow visual feedback, the ongoing simulation can optionally be rendered in 2D and 3D.
The implementation was done in C\#.\\
The level of detail of the simulation is especially high for the simulated movement behavior of the robots.
We consider the robot's momentum by emulating acceleration and deceleration behavior, collision avoidance and turning speed (see Table~\ref{tab:mu_shared_parameters}).
The emulation employs a continuous time-horizon. 
The times for activities other than robot movement, e.g. lifting or storing a pod, or picking one unit at a pick station, are constant (see Table~\ref{tab:mu_shared_parameters}).
The waypoints allow the emulated robot behavior to match real robot behaviour.
Robots that do not carry a pod can traverse underneath stored pods by using the waypoints at which the pods are stored.
Furthermore, in the buffers of the workstations, robots can take short-cuts if the buffer is (partially) empty.\\
Information about the system's state is tracked in a high level of detail, because some decision rules differ with regard to the information they require.
For example, all pods and all units on all pods are tracked exactly.
Incoming information is divided into a static and a dynamic category.
Static information includes everything describing a system instance and is completely given at start.
Static information therefore includes the number and composition of pick stations and replenishment stations, the pods, the robots, and the waypoint system used for robot navigation.
All of the decision rules proposed in this work differ in their computational complexity and therefore also in the computational time they require to reach decision.
They are, however, simple enough to be considered as ad-hoc decisions even for large system sizes.\\
In contrast to static information, the dynamic information is not completely known beforehand, but becomes available over time.
This is the case for incoming pick orders and replenishment orders submitted to the system over time by external processes.
While each replenishment order consists of a number of physical units of one SKU, each pick order consists of a set of order lines, each for one SKU, with corresponding units necessary to fulfill the line. 
We assume for both pick and replenishment orders, that there is a constant order backlog.
A constant order backlog means that when an order from the backlog is assigned to a workstation, it is immediately replaced by a newly generated order.
By keeping the order backlogs constant, we aim to analyze the system's behavior under constant pressure.
However, it also leads to the phenomenon that the system's storage space utilization (utilized space divided by total space available) in the storage area is 
affected by the performance of the decision rules controlling it, because no further virtual manager steers the process. 
E.g., if a combination of rules is replenishing quickly, the storage space utilization will increase. 
In contrast, it will decrease, if the rules are replenishing slowly.
Situations in which the storage space utilization is nearing 100\%, and only few storage places for new replenishment orders are available, lead to an 
inefficient replenishment process.
To avoid such situations, we pause replenishment order generation, if storage space utilization exceeds 85 \% and it is continued after it drops below 65 \% 
again. 
Analogously, we pause the pick order generation, if storage space utilization drops below 10 \% and resume after it exceeds 60 \% again.
The latter is done to avoid draining the inventory completely.
Since in both cases either the replenishment stations or the pick stations will become inactive due to no further orders to process, the robots will be reassigned to the remaining active stations.
This redistribution of robots across the active stations is done at any time a station becomes active or inactive, i.e. at the beginning and end of order generation pauses.\\
If a new replenishment order is received, first the rules for ROA and RPS are responsible for choosing a replenishment station and a pod (see Figure~\ref{fig:mu_problemdependencies}).
The time the decision is taken depends on the active rules.
The execution of the assignment can earliest be done as soon as there is sufficient capacity on a pod and a station available.
The commit of the assignment technically results in an insertion request (shown as red cylinders), i.e. a request that requires a robot to bring the pod to the workstation.
Multiple of these requests are then combined to an insertion task and assigned to a robot by a TA rule.
Similarly, after the POA rule selects a pick order from the backlog and the assignment is committed to a pick station, an extraction request (shown as blue cylinders) is generated, i.e. a request that requires bringing a suitable pod to the chosen station.
Up to this point, the physical units of SKUs for fulfilling the pick order are not yet chosen.
Instead, the decision is postponed and taken right before combining different requests to extraction tasks by PPS and assigning them to robots by TA.
This allows the implemented rules to exploit more information when choosing a pod for picking. 
Hence, in this work we consider PPS as a decision closely interlinked with TA.
Furthermore, the system generates store requests (shown as orange cylinders) each time a pod has to be transported to a storage location.
The PSA rule only decides the storage location for a pod that is not needed anymore and has to be returned to the storage area.
If all requests are already being handled by other robots, the robot will be assigned an idle task, thus, the robot dwells at a dwelling point until needed.
Dwelling points can be used to reduce congestion effects if there are only a few active stations compared to the number of robots, e.g. robots waiting at a storage location block others that try to pass by.
For this, the robot will park at a free storage location to avoid causing conflicts with other robots. 
The dwell point policy uses locations in the middle of the storage area to avoid blocking prominent storage locations next to the stations. 
Another type of task would be charging, which is necessary when robots run low on battery, however, in this work we assume the battery capacity to be infinite, so this type of task is ignored.
All of the tasks result in trips (shown as green cylinders), which are planned by a path planning algorithm and executed by the robots.
The only exception is when a pod can be used for another task at the same station.
The trips are planned by a PP algorithm and the resulting paths are executed by the robots.
Figure~\ref{fig:mu_problemdependencies} shows an abstract overview of these dependencies.
The exact times at which the decisions are taken depend on the respective rules, e.g. the Pod-Batch ROA rule assigns a batch of replenishment orders to the first pick station offering sufficient space while the Random ROA rule immediately assigns single replenishment orders to the first station with sufficient capacity available.
However, all of the rules have in common that they make assignments greedily while adhering to certain capacity constraints (station capacity, pod capacity, etc.).

\section{Evaluation Framework}\label{sec:mu_framework}
\begin{figure}[!htbp]
\begin{center}
	\includegraphics[width=\textwidth]{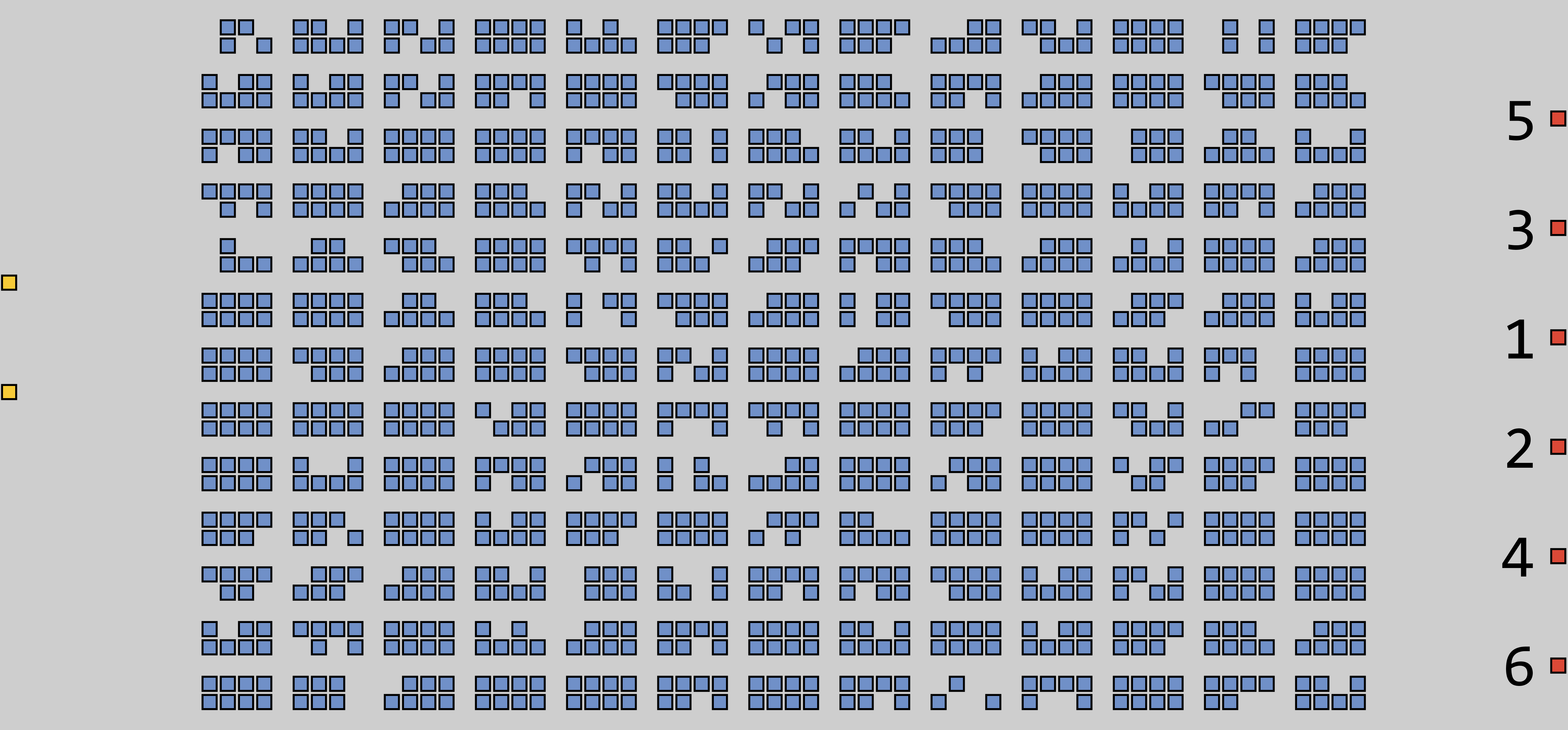}
	\caption{Top view of the layout, including pick station indices, with the storage area in the middle, replenishment stations to the left, and pick stations 
	to the right}
	\label{fig:mu_layoutsnapshot}
\end{center}
\end{figure}

This section describes the evaluation framework used to carry out the research in this paper.
Two central concepts to the evaluation framework are the Rule Configuration (RC) and the Warehouse Scenario (WS).
The RC specifies for each decision problem, which decision rule is used.
The WS specifies the warehouse layout, number of robots, number of workstations, number of SKUs, whether or not return orders are part of the operations of the 
warehouse, and pick order size.
During one simulation run the RC and WS do not change, so they can be seen as an input to a simulation run.\\
The evaluation framework consists of two phases, one varying the RCs, the other varying the WSs.
Phase 1 evaluates all 1620 possible RCs on one WS. 
For phase 1, we compare eight performance measures:
(1) unit throughput rate,
(2) pick order throughput rate,
(3) order turnover time,
(4) distance traveled per robot,
(5) order offset,
(6) fraction of orders that are late,
(7) pile-on
(8) the pick station idle time.
Unit throughput rate is the number of picked units of all SKUs per hour.
Pick order throughput rate is the number of pick orders fulfilled per hour.
Order turnover time is the average time between submitting a pick order to the backlog and fulfilling it.
Order offset is the average time between the due time and the completion time of the pick orders. 
Thus, a value smaller than zero shows how much in advance pick orders are completed. 
The rationale behind this is that follow-up processes at the distribution center are not deterministic, hence, pick orders completed earlier may improve the 
overall service level.
The pick station idle time is measured as an average across all pick stations in the system.\\
Phase 1 selects the RCs with the highest unit throughput rate.
However, among these selected best RCs, the variety in the decision rules may be low.
For a particular decision problem, all of the selected RCs may use the same decision rule.
To ensure more diversity in the RCs in phase 2, we define 6 so-called ``benchmark RCs'', see Table~\ref{tab:mu_benchmark_RCs}.
The benchmark RCs were chosen such, that all decision rules across all decision problems appear in at least one of the benchmark RCs.
Each benchmark RC has been given a name that reflects a characteristic that the decision rules have most in common.
\begin{table}[!htbp]
	\caption{Benchmark RCs}
	\label{tab:mu_benchmark_RCs}
	\small
	\setlength{\tabcolsep}{1mm}
	\begin{center}
		\resizebox{\textwidth}{!}{
		\begin{tabular}{l|rrrrr}
			\hline
			Benchmark RC	& POA 			& ROA 			& PPS 		& RPS 			& PSA \\
			\hline
			Demand 			& Due-Time	 	& Pod-Batch 	& Demand   	& Least-Demand	& Fixed \\
			Speed			& Fast-Lane		& Pod-Batch 	& Lateness 	& Emptiest		& Nearest \\
			Nearest			& FCFS 			& Random 		& Nearest 	& Nearest 		& Nearest \\
			Class 			& Common-Lines 	& Pod-Batch 	& Age		& Class 		& Class \\
			Greedy 			& Pod-Match 	& Pod-Batch 	& Pile-on 	& Emptiest		& Station-Based \\
			Random 			& Random 		& Random 		& Random 	& Random 		& Random \\
			\hline
		\end{tabular}
		}
	\end{center}
\end{table}
Phase 2 evaluates the selected RCs from phase 1 and the benchmark RCs, while varying the warehouse scenarios.
Since we are specifically interested in efficiency of RCs we neglect layout decisions for this work. 
Thus, we choose one specific layout, using the style described in Section~\ref{sec:mu_rmfs}. 
The concrete layout instance comprises 1149 pods and 1352 storage locations (~85\% filled) and is shown in Figure~\ref{fig:mu_layoutsnapshot}. 
When varying the number of pick stations during phase 2 we add workstations in the order given in Figure~\ref{fig:mu_layoutsnapshot}.

\subsection{Parameters}\label{sec:mu_parameters}

In the following we describe the used parameters in more detail. 
The parameters shared for both phases are outlined in Table~\ref{tab:mu_shared_parameters}. 
We set a continuous simulation horizon of 48 hours in order to decrease the impact of side effects like recurring replenishment overflows, which cause replenishment pauses described previously.
Within a duration of 48 hours we observe sufficient repetitions of such patterns to achieve a reasonable mitigation of these side effects.\\
Furthermore, for each RC and WS combination in phase 1 and in phase 2 we conduct 10 runs to lessen the effect of randomness.
To keep the system under continuous pressure, like described above, we keep a constant pick and replenishment order backlog of 200 orders each. 
At simulation start inventory is generated until 70 \% overall storage utilization to avoid cold starting the system. 
This is done using the same process used for generating replenishment orders during simulation and using assignment rules suiting the respective RPS rule in 
place. 
The storage capacity of a pod is set to 500 slots while the storage consumption of one SKU unit is drawn from a uniform distribution between 2 and 8 slots, 
thus, a full pod contains 100 units in average. 
The popularity of the SKUs is determined by drawing a value from an exponential distribution with parameter $\lambda=\frac{1}{2}$ for each SKU to emulate a 
typical ABC curve in e-commerce.
This popularity is the relative frequency parameter between all SKUs, thus, the frequency (if divided by the sum of all frequencies) is the probability of choosing a particular SKU when generating an order line for both replenishment and pick orders. 
One replenishment order restocks between 4 and 12 units of one SKU following a uniform distribution. To emulate due times we distinguish between priority and 
normal orders that have to be completed in 30 minutes respectively 120 minutes. 
This reflects the need for preferring important orders.\\
The movement behavior of the robots is emulated by using a maximum velocity of $1.5 \frac{m}{s}$ with acceleration and deceleration rates of $0.5 
\frac{m}{s^2}$. 
We set the rotational speed to $\frac{4}{5} \pi \frac{rad}{s}$, i.e., $2.5 s$ for a full turn.
Turning takes the same amount of time regardless of whether a robot is carrying a pod.
The time for lifting and setting down a pod is set to $3 s$. 
This should reflect the capabilities of mobile robots used in similar industry applications reasonably close. 
For the actual pick operation of one unit at a pick station we assume a constant time of $8 s$. 
The complete time for handling one unit including additional operations, like putting the product unit in the correct pick order tote, is set to $15 s$. 
This distinction is considered to allow for an early release of the robot, such that no unnecessary robot waiting times are caused. 
This is not distinguished for replenishment operations, since we assume that a robot can only leave after fully completing the put operation to the pod. 
The time of a put operation of one replenishment order is set to $20 s$.

\begin{table}[!htbp]
	\caption{Parameters shared across all simulations}
	\label{tab:mu_shared_parameters}
	\footnotesize
	\setlength{\tabcolsep}{1mm}
	\begin{center}
		\resizebox{\textwidth}{!}{
		\begin{tabular}{p{0.5\textwidth}p{0.5\textwidth}}
			\hline
			\textbf{Parameter} 											& \textbf{Value}\\
			\hline
			Simulation													& \\	
			\hline		
			Simulated duration of warehouse operations					& 48 hours\\
			Number of simulation repetitions 							& 10 repetitions\\
			Size of pick order backlog 									& 200 pick orders\\
			Size of repl. order backlog									& 200 repl. orders\\
			Layout														& 1149 pods, 1352 storage locations in $2\times4$ blocks, 12 aisles and 12 
			cross-aisles\\
			\hline
			Orders														& \\
			\hline
			Number of units per repl. order 							& uniform distribution between 4 and 12 units \\
			Amount of priority orders in pick orders					& 20 \% \\
			Priority pick order due time								& backlog submission time + 30 min. \\
			Normal pick order due time									& backlog submission time + 120 min. \\
			Threshold when pick order generation starts  				& 60\% of inventory capacity of the storage area\\
			Threshold when pick order generation stops 					& 10\% of inventory capacity of the storage area\\
			Threshold when repl. order generation starts 				& 65\% of inventory capacity of the storage area\\
			Threshold when repl. order generation stops 				& 85\% of inventory capacity of the storage area\\
			\hline
			Inventory													& \\
			\hline
			Initial inventory in the storage area						& 70\% of the inventory capacity of the storage area\\
			Space on a pod 												& 500 slots\\
			SKU frequency / popularity									& Exponential distribution, $\lambda=\frac{1}{2}$ \\
			SKU size 													& uniform distribution between 2 and 8 slots\\
			\hline
			Robot movement												& \\
			\hline			
			Robot acceleration/deceleration								& $0.5 \frac{m}{s^2}$\\
			Robot maximum velocity 										& $1.5 \frac{m}{s}$ \\
			Time needed for a full turn of a robot 						& $2.5 s$\\
			Time needed for lifting and storing a pod 					& $3 s$\\
			Time needed for picking a unit 								& $8 s$\\
			Time needed for handling a unit at pick station 			& $15 s$\\
			Time needed for putting a repl. order on a pod 				& $20 s$\\
			\hline
			Stations													& \\
			\hline			
			Repl. station capacity										& two times pod capacity\\
			Pick station capacity 										& 8 pick orders \\
			\hline
		\end{tabular}
		}
	\end{center}
\end{table}

The parameters in Table~\ref{tab:mu_shared_parameters} are shared across all conducted experiments, while the parameters in 
Table~\ref{tab:mu_phases_parameters} are depending on phase and 
scenario. 
For the first phase we assess all possible RCs for one fixed warehouse scenario. 
Note that the RPS rule Nearest and the ROA rule Pod-Batch both wait for the other one to decide first leading to no decision at all, hence, the combination of 
these rules is forbidden. 
For the fixed warehouse scenario we set the number of robots to 4 per pick station, i.e. 8 robots in the system at whole. 
Furthermore, we set the number of pick stations to 2, the number of SKUs to 1000 and exclude the processing of return orders. 
The order setting is set to Mixed. 
This means the number of lines per pick order and the number of units per order line are generated following truncated normal distributions with parameters shown in Table~\ref{tab:mu_phases_parameters}.
This is done to resemble e-commerce pick order characteristics of generally small orders with occasional larger ones in between.\\
Equation~\eqref{eq:mu_rule_combinations_equation} shows that phase 1 has 1620 RCs, and since phase 1 has 1 WS and 10 runs are conducted per RC and WS 
combination, this results in 16200 simulation runs for phase 2.
Phase 2 has 10 RCs (see Table~\ref{tab:mu_phases_parameters}) and Equation ~\eqref{eq:mu_rule_combinations_equation} shows that it has 360 WSs, which together 
with 10 runs per RC and WS combination leads to 36000 simulation runs for phase 2.
\begin{align}\nonumber
\# RC\text{ in phase 1} = \big| \left\{ ROA \right\} \times \left\{ POA \right\} \times \left\{ RPS \right\} \times \left\{ PPS \right\} \times \left\{ PSA 
\right\} \setminus & 
\\
\label{eq:mu_rule_combinations_equation}
\left\{ \left( roa, poa, rps, pps, psa \right) \mid (roa = \text{Pod-Batch}) \wedge (rps = \text{Nearest}) \right\} \big| & = 1620 
\\
\nonumber\\
\# WS\text{ in phase 2} = \big| \left\{ \text{number pick station} \right\} \times \left\{ \text{robots per pick station} \right\} & \nonumber\\ 
\times \left\{ \text{number of SKUs} \right\} \times \left\{ \text{return orders} \right\}
 \times \left\{ \text{pick order size} \right\} \big|& =\nonumber\\
6 \times 5 \times 2 \times 2 \times 3 & = 360 \label{eq:mu_ws_combinations_equation}
\end{align}
For phase 2 we limit the RCs to the 6 benchmark RCs and the 4 best ones from phase 1, i.e., the 4 RCs with highest throughput rate.
Moreover, we vary the number of pick stations from 1 through 6 and the number of robots per pick station from 2 through 6. 
This leads to a range from 2 robots in the system to 36 robots across all WSs. 
In addition to WSs with 1000 SKU, we also assess WSs with 10000 SKUs stored in the system. 
For the order size we define two additional settings of small and large orders. 
For the Small pick order size, only single line / single unit pick orders are generated. 
For the Large pick order size, the distributions from the Mixed order setting are used but the $\text{min}$ parameter for both is set to 2. 
Lastly, in WSs where we emulate the processing of return orders, 30 \% of the generated replenishment orders are single unit. 
The total number of RC and WS combinations for the phase 2 is therefore 3600, which leads to 36000 simulation runs.

\begin{table}[!htb]
	\small
	\setlength{\tabcolsep}{1mm}
	\caption{Varied parameters for phase 1 and 2}
	\label{tab:mu_phases_parameters}
	\begin{center}
		\resizebox{\textwidth}{!}{
		\begin{tabular}{L{0.26\textwidth}|R{0.35\textwidth}|R{0.35\textwidth}}
			\hline
			Parameter & Phase 1 values & Phase 2 values \\
			\hline
			Rule configurations (RCs) & 1620 RCs & 6 Benchmark RCs \\
			& & + 4 best RCs from phase 1 \\
			\hline
			Number of pick stations & 2 & 1, 2, 3, 4, 5, 6 \\
			Robots per pick station & 4 & 2, 3, 4, 5, 6 \\
			Number of SKUs & 1000 & 1000, 10000 \\
			Return orders & 0 \% & 0 \%, 30 \% \\
			\hline
			Pick order size & \textit{Mixed} - line \& unit distributions: \newline $\mu=1,\sigma=1, \text{min} = 1, 
			\text{max} = 4$ \newline $\mu=1,\sigma=0.3, \text{min} = 1, \text{max} = 3$ & \textit{Small} - line \& unit distributions: \newline $\text{min} = 1, 
			\text{max} = 1$ \newline $\text{min} = 1, \text{max} = 1$ \newline \textit{Mixed} - line \& unit distributions: \newline $\mu=1,\sigma=1, \text{min} = 1, 
			\text{max} = 4$ \newline $\mu=1,\sigma=0.3, \text{min} = 1, \text{max} = 3$ \newline \textit{Large} - line \& unit distributions: \newline $\mu=1,\sigma=1, \text{min} = 2, 
			\text{max} = 4$ \newline $\mu=1,\sigma=0.3, \text{min} = 2, \text{max} = 3$ \\
			\hline
			\# RC & 1620 & 10\\
			\# WS & 1 & 360\\
			\# RC$\times$WS & 1620 & 3600\\
			\# simulation runs & 16200 & 36000\\
			\hline
		\end{tabular}
		}
	\end{center}
\end{table}

\section{Computational Results}\label{sec:mu_results}
This section shows the results from phase 1 and phase 2 of the evaluation framework.
Throughout this section, the unit throughput rate is presented as a percentage of the upper bound on the unit throughput rate.
The unit throughput rate is presented in this way to facilitate interpretation and comparison of results across experiments.
Moreover, the RMFS is supposed to have high pick rates as it eliminates the need of walking for the workers, while the robots are supposed to supply the pickers with a constant stream of pods to pick from.
Presenting the unit throughput rate as a percentage shows clearly to what extent these aims are achieved.
The upper bound is discussed in more detail in Appendix~\ref{sec:mu_upperbound}.
The length of the confidence intervals is always less than 1\% of the mean, based on 10 runs per RC and WS combination, and therefore does not add much information.

\subsection{Phase 1}

\begin{table}[htb]
\caption{Correlations between the different performance measures for first phase}
\label{tab:mu_correlations}
\renewcommand{\arraystretch}{1}
\centering
\small
\setlength{\tabcolsep}{0.5mm}
\resizebox{\textwidth}{!}{
\begin{tabular}{l|rr|cccccccc}
& $\mu$ & $\sigma$ & \rotatebox{90}{Unit throughput} & \rotatebox{90}{Order throughput} & \rotatebox{90}{Order turnover time} & \rotatebox{90}{Distance traveled} & \rotatebox{90}{Order offset} & \rotatebox{90}{Late orders} & \rotatebox{90}{Pile-on} & \rotatebox{90}{Station idle time}\\
\hline
Unit throughput & 0.556 & 0.189  & - & - & - & - & - & - & - & -\\
Order throughput & 241.963 & 82.234 & \cellcolor{mediumgreen!99.800!mediumyellow!80!white}1.000 & - & - & - & - & - & - & -\\
Order turnover time & 3549.625 & 1220.445 & \cellcolor{mediumyellow!4.974!mediumred!80!white}-0.950 & \cellcolor{mediumyellow!4.969!mediumred!80!white}-0.950 & 
- & - & - & - & - & -\\
Distance traveled & 122598.768 & 19433.860 & \cellcolor{mediumyellow!4.765!mediumred!80!white}-0.952 & \cellcolor{mediumyellow!4.751!mediumred!80!white}-0.952 & \cellcolor{mediumgreen!87.987!mediumyellow!80!white}0.880 & - & - & - & - & -\\
Order offset & -2565.458 & 1224.985 & \cellcolor{mediumyellow!4.992!mediumred!80!white}-0.950 & \cellcolor{mediumyellow!4.988!mediumred!80!white}-0.950 & \cellcolor{mediumgreen!99.800!mediumyellow!80!white}1.000 & \cellcolor{mediumgreen!87.961!mediumyellow!80!white}0.880 & - & - & - & -\\
Late orders & 0.187 & 0.115 & \cellcolor{mediumyellow!40.957!mediumred!80!white}-0.590 & \cellcolor{mediumyellow!40.948!mediumred!80!white}-0.591 & \cellcolor{mediumgreen!68.516!mediumyellow!80!white}0.685 & \cellcolor{mediumgreen!54.898!mediumyellow!80!white}0.549 & \cellcolor{mediumgreen!68.357!mediumyellow!80!white}0.684 & - & - & -\\
Pile-on & 2.438 & 1.450 & \cellcolor{mediumgreen!89.938!mediumyellow!80!white}0.899 & \cellcolor{mediumgreen!89.942!mediumyellow!80!white}0.899 & \cellcolor{mediumyellow!19.834!mediumred!80!white}-0.802 & \cellcolor{mediumyellow!20.432!mediumred!80!white}-0.796 & \cellcolor{mediumyellow!19.829!mediumred!80!white}-0.802 & \cellcolor{mediumyellow!55.214!mediumred!80!white}-0.448 & - & -\\
Station idle time & 0.450 & 0.186 & \cellcolor{mediumyellow!0.200!mediumred!80!white}-1.000 & \cellcolor{mediumyellow!0.200!mediumred!80!white}-1.000 & \cellcolor{mediumgreen!95.041!mediumyellow!80!white}0.950 & \cellcolor{mediumgreen!95.241!mediumyellow!80!white}0.952 & \cellcolor{mediumgreen!95.022!mediumyellow!80!white}0.950 & \cellcolor{mediumgreen!59.063!mediumyellow!80!white}0.591 & \cellcolor{mediumyellow!10.090!mediumred!80!white}-0.899 & -\\
\end{tabular}
}
\end{table}

\begin{figure}[htb]
	\centering
	\includegraphics[width = 0.98\textwidth]{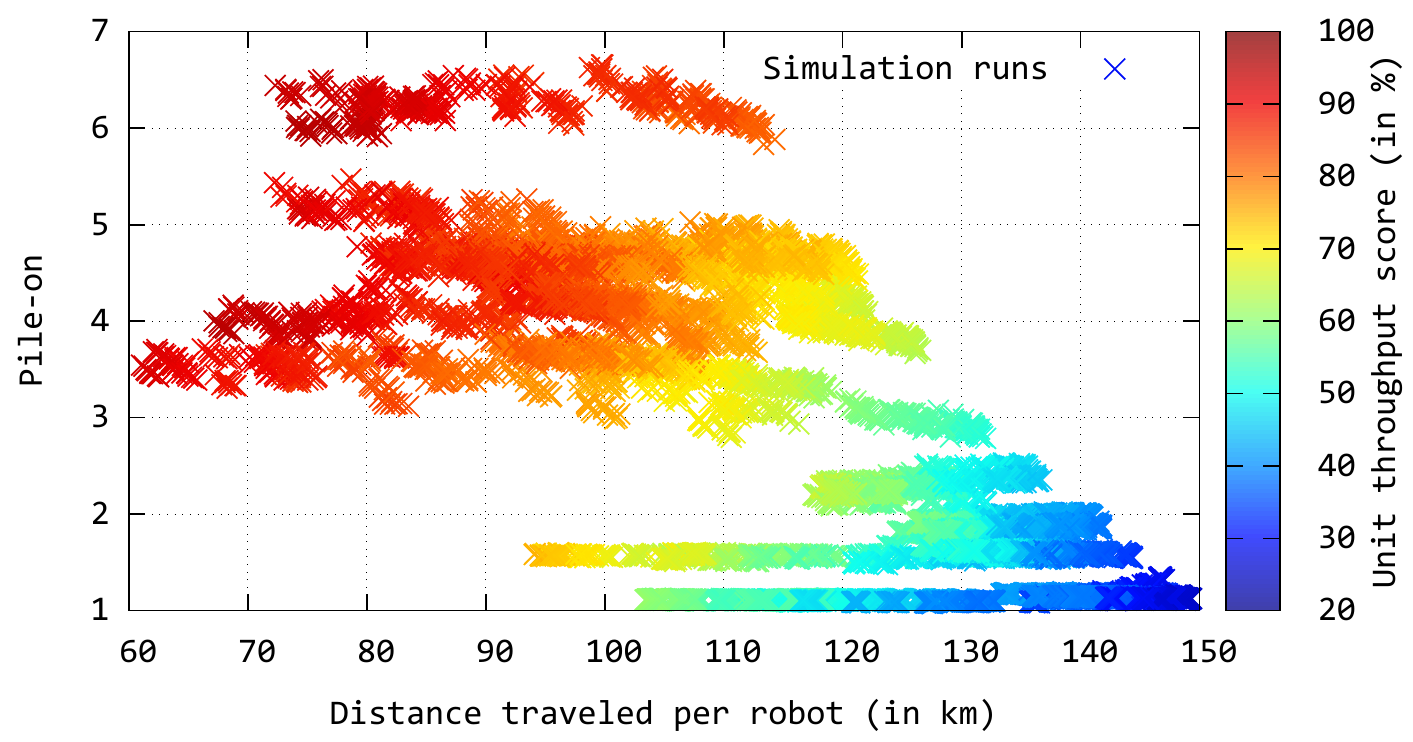}
	\caption{Scatter plot for pile-on vs. traveled distance per robot colored by the achieved throughput rate score for all simulation runs of the first phase}
	\label{fig:mu_botdistancepileonthroughputscatterplot}
\end{figure}

The first phase aims to investigate throughput performance and the impact per decision problem of decision rules on throughput.
Furthermore, we assess the behavior of the different output measures depending on decision rule selection.
For this, Table~\ref{tab:mu_correlations} shows how across these simulations the seven previously introduced performance measures correlate with each other.
At first, we can observe that as the unit throughput rate score improves, the other performance measures improve as well.
As the unit throughput rate score increases, pick order throughput rate and pile-on increase as well, whereas the order turnover time, the distance that robots travel, the order offset, the fraction of orders that miss their due time, and the station idle time decreases.
Although it is not clear what the exact causal relationships are, the correlations suggest that pile-on and the distance traveled by the robots are the main drivers behind these improvements.
With higher pile-on, more units are picked per pod, so order lines are fulfilled more quickly and fewer trips are needed to fulfill the pick orders.
This also causes longer processing times for each pod at the pick station, which in turn increases the time for the next robot to queue and become ready at the station.
In other words: a more continuous input of inventory at the pick station is achieved.
Additionally, fewer trips for the pick process free up robots to do more replenishment tasks.
With less distance traveled by the robots we expect pods to be presented at the pick stations more continuously.
Similar to the pile-on this effect enables more continuous picking, which in turn increases the overall unit throughput rate.
Both measures, pile-on and the traveled distance, are intermediate measures affected by the choice of strategy for the different decision problems, i.e., a better score in both helps decreasing the idle time at the stations, which in turn helps increasing the throughput.
An increased throughput, in the constant pick order backlog setting of this work, also decreases the turnover time of pick orders and the due time offset.
Only the number of orders being late is not strongly correlated with the two main throughput drivers.
The two main throughput drivers can also be observed when looking at a scatter plot of all simulation runs of the first phase (see 
Figure~\ref{fig:mu_botdistancepileonthroughputscatterplot}).
Here we can see the best results in unit throughput rate score are achieved with a high pile-on and less distance traveled per robot.
The group of simulation runs with least distance traveled per bot and a pile-on around 4 are RCs involving the Nearest PPS rule, while the simulation runs with highest pile-on (greater 5) at the top of the plot are RCs involving the Demand PPS rule.
In both groups we find runs with the highest unit throughput rate score, hence, a higher throughput is not only achieved by a high pile-on.
In particular within the top ten RCs in terms of unit throughput rate score the pile-on ranges between 3.84 and 6.36, while the distance traveled per bot ranges between 68.04 km to 80.36 km.
Hence, pile-on and the traveled distance enable higher throughput, but may also compensate for each other.
This is particularly interesting, because both come at operational costs.
For traveled distance this is energy consumption and robot wear, while for pile-on it may be costs arising from potentially more complex replenishment processes.
Furthermore, within both groups better results are obtained with RCs also involving the Pod-Match POA rule, which causes an additional boost in pile-on.

In Figure~\ref{fig:mu_botdistancepileonthroughputscatterplot} we also observe a 'cutoff' of simulation runs in the upper right and bottom left areas.
This can be explained by the longer handling time at the station resulting from a higher pile-on.
I.e., the longer a robot needs to wait at a station for the picking to finish the less it can travel in the meantime.
Thus, rules increasing pile-on may help reducing the necessary travel distance, and by this also robot wear and energy consumption.

The pick order throughput rate is neglected completely in the remainder of this work, because it almost completely aligns with the unit throughput rate score. 
The reason for this is the constant backlog of 200 pick orders over 48 hours:
with a pick order throughput rate of 241.963 completed orders per hour in average, omitting certain pick orders is almost impossible. 
Hence, we cannot observe a potential temporary throughput gain by preferring smaller or larger orders.
In order to investigate the trade-off between picking many units and completing more pick orders an experiment with a fixed set of backlogged pick orders over a fixed period of time should be devised. 
For this, the possibly tedious processing of leftover pick orders, which are presumably harder to pick quickly, needs to be investigated.
We leave this work for future research.

\begin{table}[!htbp]
	\caption{Average unit throughput rates as percentages of the upper bound for all rules, together with the \textbf{best} / \textbf{worst} performance multiplier per decision problem}
	\label{tab:mu_multiplier}
	\setlength{\tabcolsep}{0.5mm}
	\small
	\centering
	\resizebox{\textwidth}{!}{
	\begin{tabular}{l|rrrrrr|r}
		& & & & & & & Mult. $\left(\frac{\text{best}}{\text{worst}}\right)$ \\
		\hline
		POA & Common-Lines & Due-Time & Fast-Lane & FCFS & Pod-Match & Random & \\
		 & 50.93\% & 41.93\% & 76.13\% & 41.81\% & \textbf{81.18\%} & \textbf{41.71\%} & 1.946 \\
		\hline
		ROA & Random & Pod-Batch & & & & & \\
		 & \textbf{53.71\%} & \textbf{57.99\%} &  &  &  &  & 1.080 \\
		\hline
		PPS & Age & Demand & Lateness & Nearest & Pile-on & Random & \\
		 & 61.50\% & 52.70\% & \textbf{48.63\%} & \textbf{62.16\%} & 59.82\% & 48.88\% & 1.278 \\
		\hline
		RPS & Class & Nearest & Emptiest & Least-Demand & Random & & \\
		 & 56.16\% & 58.42\% & \textbf{59.63\%} & 57.71\% & \textbf{47.56\%} &  & 1.254 \\
		\hline
		PSA & Class & Fixed & Nearest & Random & Station-Based & & \\
		 & 55.91\% & 54.08\% & \textbf{58.79\%} & \textbf{53.60\%} & 55.70\% &  & 1.097 \\
	\end{tabular}
	}
\end{table}

Table~\ref{tab:mu_multiplier} shows for each decision problem the unit throughput rate score for each of the decision rules, averaged across all simulations in phase 1.
We calculate the multiplier by dividing the highest unit throughput rate by the lowest.
As the multiplier in unit throughput rates is rather large for the POA decision problem, system integrators and RMFS suppliers may benefit from carefully selecting a POA decision rule and from investigating better decision rules for this decision problem.
The multiplier for the Replenishment Order Assignment is near 1, indicating that using a different decision rule does not offer much performance improvements.
However, we note that we keep the sequence of incoming replenishment orders fixed at all times in this work, which limits improvement potential. Nevertheless, we expect limited degrees-of-freedom in replenishment operations to be more realistic, because the sequence will typically be a result of preceding operations or systems.
Moreover, the limited number of replenishment stations diminishes the impact of ROA decision rules even more.
Furthermore, the impact of the Pod Storage Assignment selection rule seems to be fairly low. This may be a reason of the quite small layout. We expect the impact of PSA decision rules to increase with the size of the instance layout, because the effect on the traveled distance would grow by a large amount.

\begin{figure}[htb]
\caption{Unit throughput rate performance of all runs grouped per rule}
\label{fig:mu_boxplots}
\centering
\begin{subfigure}[b]{0.49\textwidth}
	\includegraphics[width = \textwidth]{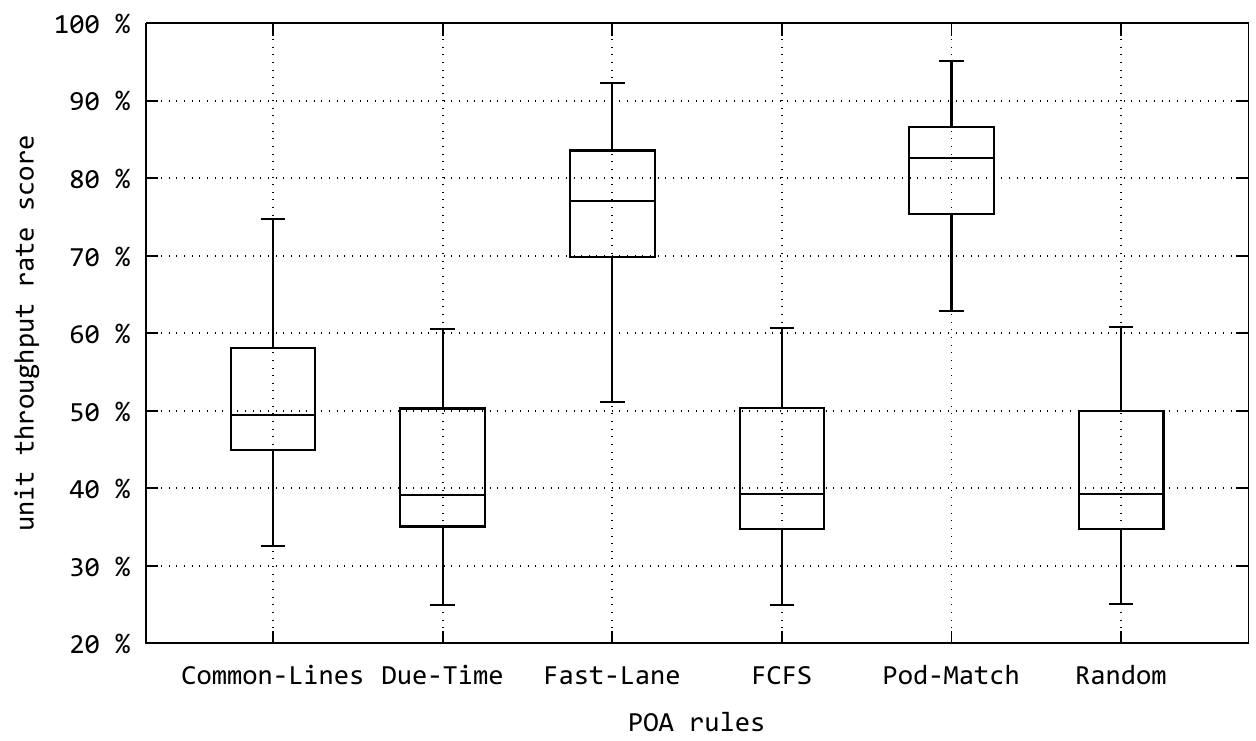}
	\label{fig:mu_boxplotpoa}
\end{subfigure}
\begin{subfigure}[b]{0.49\textwidth}
	\includegraphics[width = \textwidth]{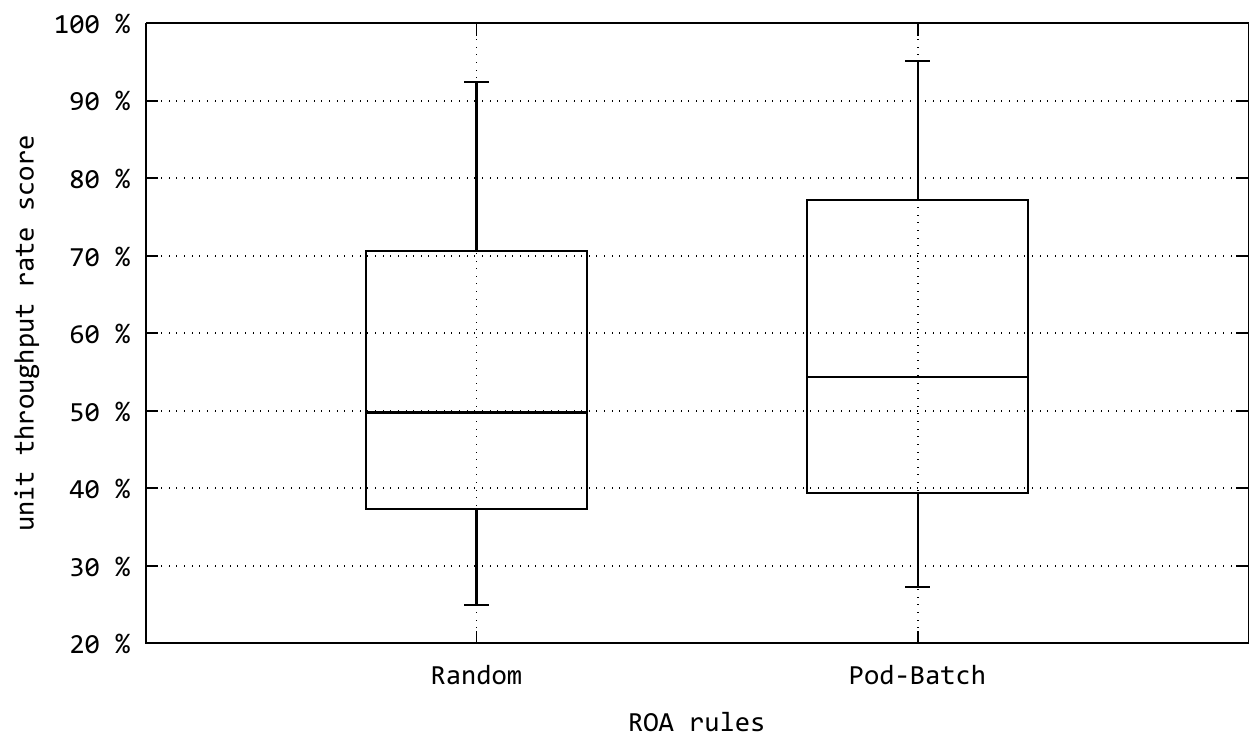}
	\label{fig:mu_boxplotroa}
\end{subfigure}
\\[-4mm]
\begin{subfigure}[b]{0.49\textwidth}
	\includegraphics[width = \textwidth]{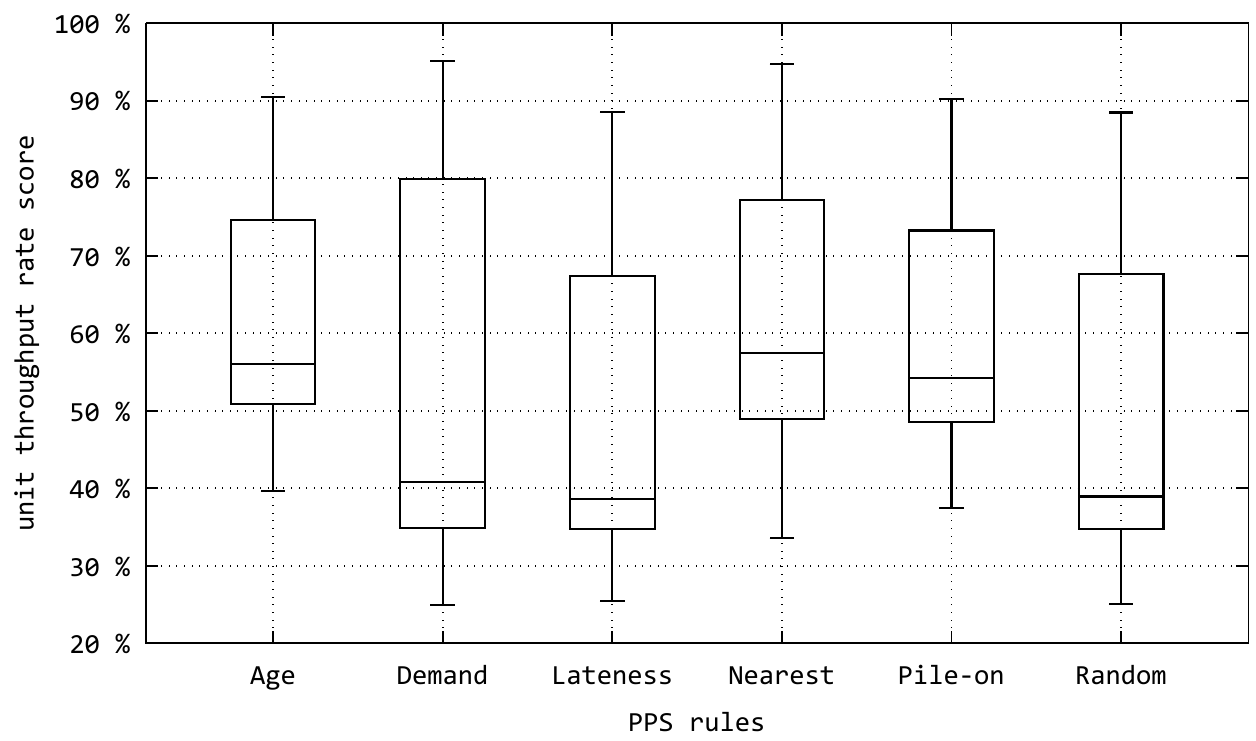}
	\label{fig:mu_boxplotpps}
\end{subfigure}
\begin{subfigure}[b]{0.49\textwidth}
	\includegraphics[width = \textwidth]{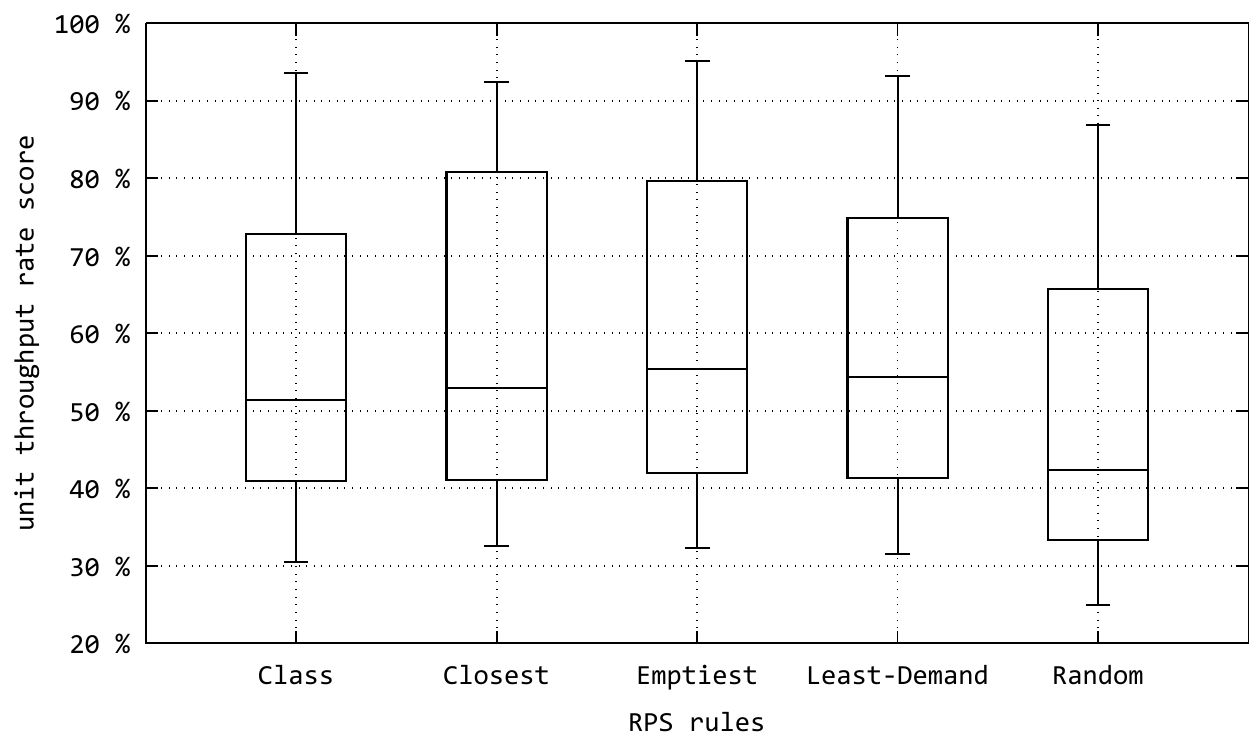}
	\label{fig:mu_boxplotrps}
\end{subfigure}
\\[-4mm]
\begin{subfigure}[b]{0.49\textwidth}
	\includegraphics[width = \textwidth]{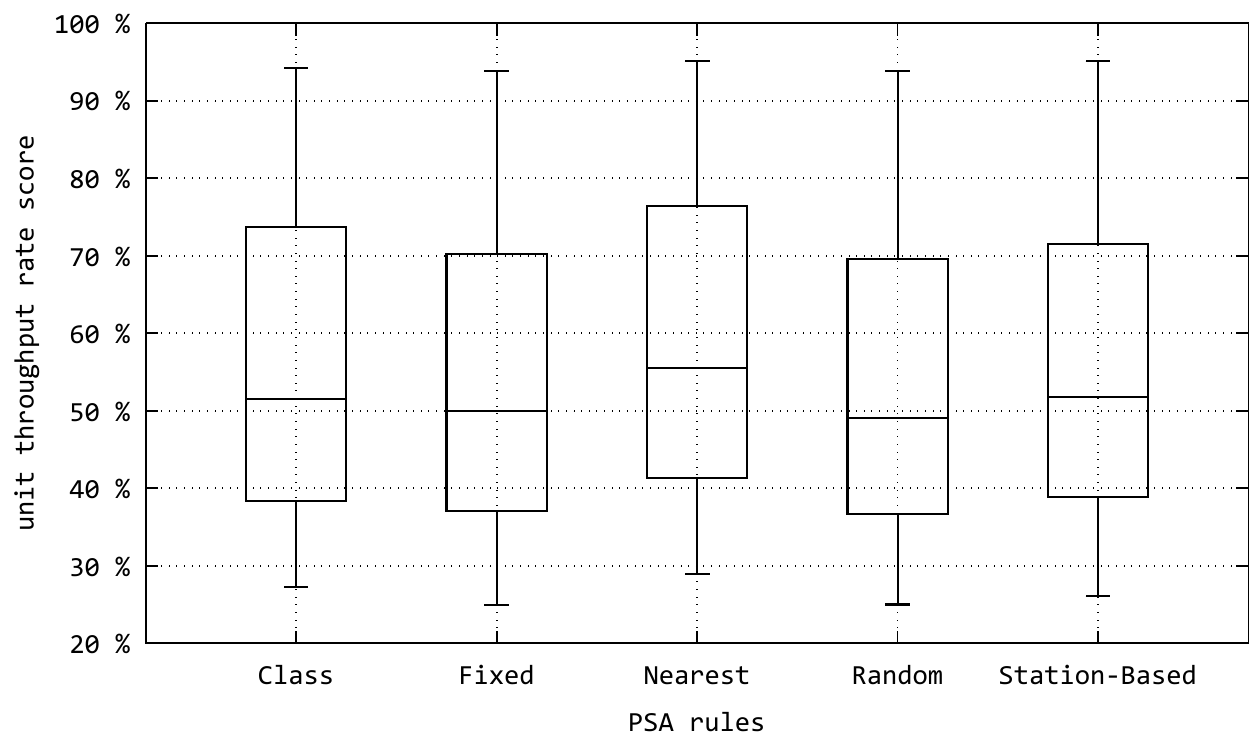}
	\label{fig:mu_boxplotpsa}
\end{subfigure}
\end{figure}

In the following we analyze the achieved throughput performance per decision rule.
For this, Figure~\ref{fig:mu_boxplots} shows the box-plots of unit throughput rate scores for each decision problem grouped per decision rule.
The boundaries of the boxes are determined by the upper and lower quartile while the line in the middle indicates the median value.
The whiskers extend from the boxes to the minimum and maximum values.
The first observation is that throughput performance of the RMFS is most sensitive to the choice of POA decision rule among the defined decision rules.
This aligns with the previously observed correlations, because the choice of POA immediately affects the pile-on, which is identified as a major performance driver.
The best performing POA strategies are FastLane and PodMatch, which both look at the incoming pods at a pick station when assigning new pick orders from the backlog.
This suggests that a strategy aligning pick orders with the content of incoming pods seems most promising for throughput efficiency.
This backs up the findings of \cite{A810}.
Although the Common-Lines rule exploits a similar greedy strategy, it achieves substantially less throughput.
Hence, only matching pick orders to each other but not to the content of the pods squanders throughput capabilities of the system.
All other POA decision rules achieve similar throughput performance, since they do not consider order characteristics that would affect pile-on or traveled distance.\\
When looking at the PPS rule box-plots the average best throughput performance with least variance is achieved by the Age, Nearest and Pile-on rules. 
All of them focus either on maximizing the pile-on or minimizing the traveled distance. 
Although the Age rule does only indirectly maximize pile-on, it achieves a higher average pile-on of 2.92 among all RCs containing it than the actual Pile-on 
rule, which achieves an average pile-on of 2.79. 
The Demand rule has the highest spread across PPS rules with a very low median, but also provides some top performing RCs (see Table~\ref{tab:mu_best_RCs}). 
This suggests that the throughput performance of the rule has a higher dependency on the selection of other rules.\\
Although the variation among the ROA decision rules is small, we observe a slightly better throughput performance by the Pod-Batch rule. 
This is a reason of the smaller number of trips necessary when batching replenishment orders.\\
Many of the top performing RCs contain the Emptiest or Closest RPS decision rule. 
The main reason for the good throughput performance again seems to rely on fewer and shorter trips. 
The Emptiest rule decreases the number of trips, because more replenishment orders are stored in pods at once until it is full. 
E.g., only 31.03 \% pods need to be brought to replenishment stations in average when compared to the Random rule. 
The Closest rule benefits from a similar effect since the same (closest) pod is used for further replenishment orders even while it is already approaching. 
Furthermore, Closest decreases the distance per replenishment trip, because nearer pods are used. 
The Random rule performs worst for RPS. The main reason for this is that too many trips are caused by randomly selecting pods while only storing few 
replenishment orders per trip.\\
Among the PSA decision rules we observe the best throughput performance for the Nearest strategy. 
This is again mainly caused by the shorter trips for the robots. 
When comparing the Nearest and the Station-based rule we see an overall benefit when ... shorter trips for replenishment operations too, and not using the 
robots working for replenishment operations to prepare pods for pick operations. 
However, this depends on the queue length at stations and the distribution of robots between replenishment and picking. 
I.e., if longer queue times are expected at replenishment stations, moving pods nearer to the pick stations when returning them to the inventory may improve 
overall throughput performance. 
The Fixed and Random decision rules differ little in their performance. 
The main reason for this is that the storage location per pod in the Fixed rule is randomly selected. 
Thus, leading to a very similar behavior.\\
Due to the large sample sizes, the results of ANOVA and Tukey's range tests rejected the hypotheses that the means were equals at the 0.05 significance level within groups and pair-wise, with five exceptions. The null hypothesis of equal means was not rejected at the 0.05 significance level for POA rules FCFS and Due-Time, for Random and Due-Time, and for Random and FCFS. Furthermore, for PPS rules Random and Lateness the hypothesis of equal means could not be rejected, and for PSA rules Station-Based and Class.

\subsection{Phase 2}

\begin{table}[!htbp]
	\caption{RCs with best throughput score selected from first phase (performance is unit throughput rate score)}
	\label{tab:mu_best_RCs}
	\small
	\centering
	\setlength{\tabcolsep}{1mm}
	\resizebox{\textwidth}{!}{
	\begin{tabular}{l|rrrrr|r}
		RC rank & POA & ROA & PPS & RPS & PSA & performance \\
		\hline
		1 & Pod-Match & Pod-Batch & Demand & Emptiest & Nearest & 94.81 \% \\
		2 & Pod-Match & Pod-Batch & Demand & Emptiest & Station-Based & 94.63 \% \\
		3 & Pod-Match & Pod-Batch & Nearest & Emptiest & Nearest & 94.43 \% \\
		4 & Pod-Match & Pod-Batch & Demand & Emptiest & Class & 94.00 \% \\
	\end{tabular}
	}
\end{table}

From the 1620 RCs in phase 1, the four with the highest unit throughput rate (see Table~\ref{tab:mu_best_RCs}) together with the benchmark RCs form the set of 
ten RCs used in phase 2.
The main purpose of phase 2 is to examine how well the RCs perform under different circumstances.
In the following we analyze the results obtained for the 12 warehouse scenarios and 30 resource settings described before (see 
Section~\ref{sec:mu_parameters}).

\begin{table}[htb]
\caption{Best unit throughput rate score for all scenarios, robots per pick station and numbers of pick stations. Scenario abbreviations: [SKU count: 1000 (1K), 10000 (10K)]-[Order size: Small (S), Medium (M), Large (L)]-[Return orders: yes (R), no (N)]}
\label{tab:mu_ITRSallscenarios}
\centering
\setlength{\tabcolsep}{0.25mm}
\resizebox{\textwidth}{!}{
\begin{tabular}{l|rrrrr|rrrrr|rrrrr|rrrrr|rrrrr|rrrrr}
Stations & \multicolumn{5}{c|}{1} & \multicolumn{5}{c|}{2} & \multicolumn{5}{c|}{3} & \multicolumn{5}{c|}{4} & \multicolumn{5}{c|}{5} & \multicolumn{5}{c}{6} \\
Robots & 2 & 3 & 4 & 5 & 6 & 2 & 3 & 4 & 5 & 6 & 2 & 3 & 4 & 5 & 6 & 2 & 3 & 4 & 5 & 6 & 2 & 3 & 4 & 5 & 6 & 2 & 3 & 4 & 5 & 6 \\
\hline
 1K-S-N & \cellcolor{mediumyellow!59!mediumred!80!white}44 & \cellcolor{mediumgreen!56!mediumyellow!80!white}82 & \cellcolor{mediumgreen!80!mediumyellow!80!white}91 & \cellcolor{mediumgreen!93!mediumyellow!80!white}97 & \cellcolor{mediumgreen!95!mediumyellow!80!white}97 & \cellcolor{mediumyellow!98!mediumred!80!white}59 & \cellcolor{mediumgreen!73!mediumyellow!80!white}89 & \cellcolor{mediumgreen!88!mediumyellow!80!white}94 & \cellcolor{mediumgreen!94!mediumyellow!80!white}97 & \cellcolor{mediumgreen!95!mediumyellow!80!white}98 & \cellcolor{mediumgreen!10!mediumyellow!80!white}64 & \cellcolor{mediumgreen!75!mediumyellow!80!white}90 & \cellcolor{mediumgreen!88!mediumyellow!80!white}95 & \cellcolor{mediumgreen!94!mediumyellow!80!white}97 & \cellcolor{mediumgreen!96!mediumyellow!80!white}98 & \cellcolor{mediumgreen!1!mediumyellow!80!white}60 & \cellcolor{mediumgreen!69!mediumyellow!80!white}87 & \cellcolor{mediumgreen!83!mediumyellow!80!white}93 & \cellcolor{mediumgreen!94!mediumyellow!80!white}97 & \cellcolor{mediumgreen!96!mediumyellow!80!white}98 & \cellcolor{mediumyellow!92!mediumred!80!white}57 & \cellcolor{mediumgreen!69!mediumyellow!80!white}87 & \cellcolor{mediumgreen!85!mediumyellow!80!white}93 & \cellcolor{mediumgreen!94!mediumyellow!80!white}97 & \cellcolor{mediumgreen!97!mediumyellow!80!white}98 & \cellcolor{mediumyellow!98!mediumred!80!white}59 & \cellcolor{mediumgreen!70!mediumyellow!80!white}87 & \cellcolor{mediumgreen!85!mediumyellow!80!white}94 & \cellcolor{mediumgreen!95!mediumyellow!80!white}97 & \cellcolor{mediumgreen!97!mediumyellow!80!white}98 \\
 1K-S-R & \cellcolor{mediumyellow!63!mediumred!80!white}46 & \cellcolor{mediumgreen!55!mediumyellow!80!white}82 & \cellcolor{mediumgreen!82!mediumyellow!80!white}92 & \cellcolor{mediumgreen!94!mediumyellow!80!white}97 & \cellcolor{mediumgreen!95!mediumyellow!80!white}97 & \cellcolor{mediumyellow!98!mediumred!80!white}59 & \cellcolor{mediumgreen!73!mediumyellow!80!white}89 & \cellcolor{mediumgreen!87!mediumyellow!80!white}94 & \cellcolor{mediumgreen!95!mediumyellow!80!white}97 & \cellcolor{mediumgreen!96!mediumyellow!80!white}98 & \cellcolor{mediumgreen!8!mediumyellow!80!white}63 & \cellcolor{mediumgreen!77!mediumyellow!80!white}90 & \cellcolor{mediumgreen!90!mediumyellow!80!white}95 & \cellcolor{mediumgreen!95!mediumyellow!80!white}97 & \cellcolor{mediumgreen!96!mediumyellow!80!white}98 & \cellcolor{mediumgreen!4!mediumyellow!80!white}61 & \cellcolor{mediumgreen!72!mediumyellow!80!white}88 & \cellcolor{mediumgreen!84!mediumyellow!80!white}93 & \cellcolor{mediumgreen!95!mediumyellow!80!white}97 & \cellcolor{mediumgreen!96!mediumyellow!80!white}98 & \cellcolor{mediumyellow!90!mediumred!80!white}56 & \cellcolor{mediumgreen!69!mediumyellow!80!white}87 & \cellcolor{mediumgreen!84!mediumyellow!80!white}93 & \cellcolor{mediumgreen!94!mediumyellow!80!white}97 & \cellcolor{mediumgreen!97!mediumyellow!80!white}98 & \cellcolor{mediumyellow!88!mediumred!80!white}55 & \cellcolor{mediumgreen!67!mediumyellow!80!white}86 & \cellcolor{mediumgreen!85!mediumyellow!80!white}93 & \cellcolor{mediumgreen!95!mediumyellow!80!white}97 & \cellcolor{mediumgreen!97!mediumyellow!80!white}98 \\
 1K-M-N & \cellcolor{mediumyellow!62!mediumred!80!white}45 & \cellcolor{mediumgreen!57!mediumyellow!80!white}83 & \cellcolor{mediumgreen!81!mediumyellow!80!white}92 & \cellcolor{mediumgreen!94!mediumyellow!80!white}97 & \cellcolor{mediumgreen!95!mediumyellow!80!white}98 & \cellcolor{mediumgreen!1!mediumyellow!80!white}60 & \cellcolor{mediumgreen!75!mediumyellow!80!white}90 & \cellcolor{mediumgreen!89!mediumyellow!80!white}95 & \cellcolor{mediumgreen!95!mediumyellow!80!white}98 & \cellcolor{mediumgreen!96!mediumyellow!80!white}98 & \cellcolor{mediumgreen!10!mediumyellow!80!white}64 & \cellcolor{mediumgreen!77!mediumyellow!80!white}90 & \cellcolor{mediumgreen!90!mediumyellow!80!white}95 & \cellcolor{mediumgreen!95!mediumyellow!80!white}98 & \cellcolor{mediumgreen!97!mediumyellow!80!white}98 & \cellcolor{mediumgreen!1!mediumyellow!80!white}60 & \cellcolor{mediumgreen!72!mediumyellow!80!white}88 & \cellcolor{mediumgreen!85!mediumyellow!80!white}93 & \cellcolor{mediumgreen!95!mediumyellow!80!white}98 & \cellcolor{mediumgreen!97!mediumyellow!80!white}98 & \cellcolor{mediumyellow!92!mediumred!80!white}57 & \cellcolor{mediumgreen!71!mediumyellow!80!white}88 & \cellcolor{mediumgreen!86!mediumyellow!80!white}94 & \cellcolor{mediumgreen!96!mediumyellow!80!white}98 & \cellcolor{mediumgreen!97!mediumyellow!80!white}98 & \cellcolor{mediumyellow!98!mediumred!80!white}59 & \cellcolor{mediumgreen!71!mediumyellow!80!white}88 & \cellcolor{mediumgreen!87!mediumyellow!80!white}94 & \cellcolor{mediumgreen!96!mediumyellow!80!white}98 & \cellcolor{mediumgreen!97!mediumyellow!80!white}98 \\
 1K-M-R & \cellcolor{mediumyellow!62!mediumred!80!white}45 & \cellcolor{mediumgreen!55!mediumyellow!80!white}82 & \cellcolor{mediumgreen!82!mediumyellow!80!white}92 & \cellcolor{mediumgreen!95!mediumyellow!80!white}97 & \cellcolor{mediumgreen!96!mediumyellow!80!white}98 & \cellcolor{mediumyellow!98!mediumred!80!white}59 & \cellcolor{mediumgreen!75!mediumyellow!80!white}89 & \cellcolor{mediumgreen!89!mediumyellow!80!white}95 & \cellcolor{mediumgreen!96!mediumyellow!80!white}98 & \cellcolor{mediumgreen!96!mediumyellow!80!white}98 & \cellcolor{mediumgreen!8!mediumyellow!80!white}63 & \cellcolor{mediumgreen!79!mediumyellow!80!white}91 & \cellcolor{mediumgreen!91!mediumyellow!80!white}96 & \cellcolor{mediumgreen!96!mediumyellow!80!white}98 & \cellcolor{mediumgreen!97!mediumyellow!80!white}98 & \cellcolor{mediumgreen!4!mediumyellow!80!white}62 & \cellcolor{mediumgreen!73!mediumyellow!80!white}89 & \cellcolor{mediumgreen!85!mediumyellow!80!white}93 & \cellcolor{mediumgreen!96!mediumyellow!80!white}98 & \cellcolor{mediumgreen!97!mediumyellow!80!white}98 & \cellcolor{mediumyellow!90!mediumred!80!white}56 & \cellcolor{mediumgreen!70!mediumyellow!80!white}88 & \cellcolor{mediumgreen!86!mediumyellow!80!white}94 & \cellcolor{mediumgreen!96!mediumyellow!80!white}98 & \cellcolor{mediumgreen!97!mediumyellow!80!white}98 & \cellcolor{mediumyellow!89!mediumred!80!white}56 & \cellcolor{mediumgreen!68!mediumyellow!80!white}87 & \cellcolor{mediumgreen!86!mediumyellow!80!white}94 & \cellcolor{mediumgreen!96!mediumyellow!80!white}98 & \cellcolor{mediumgreen!98!mediumyellow!80!white}98 \\
 1K-L-N & \cellcolor{mediumyellow!86!mediumred!80!white}54 & \cellcolor{mediumgreen!59!mediumyellow!80!white}83 & \cellcolor{mediumgreen!85!mediumyellow!80!white}93 & \cellcolor{mediumgreen!99!mediumyellow!80!white}99 & \cellcolor{mediumgreen!100!mediumyellow!80!white}99 & \cellcolor{mediumgreen!14!mediumyellow!80!white}66 & \cellcolor{mediumgreen!75!mediumyellow!80!white}90 & \cellcolor{mediumgreen!93!mediumyellow!80!white}97 & \cellcolor{mediumgreen!99!mediumyellow!80!white}99 & \cellcolor{mediumgreen!100!mediumyellow!80!white}99 & \cellcolor{mediumgreen!20!mediumyellow!80!white}68 & \cellcolor{mediumgreen!78!mediumyellow!80!white}91 & \cellcolor{mediumgreen!92!mediumyellow!80!white}96 & \cellcolor{mediumgreen!99!mediumyellow!80!white}99 & \cellcolor{mediumgreen!100!mediumyellow!80!white}99 & \cellcolor{mediumgreen!18!mediumyellow!80!white}67 & \cellcolor{mediumgreen!71!mediumyellow!80!white}88 & \cellcolor{mediumgreen!87!mediumyellow!80!white}94 & \cellcolor{mediumgreen!98!mediumyellow!80!white}99 & \cellcolor{mediumgreen!100!mediumyellow!80!white}99 & \cellcolor{mediumgreen!7!mediumyellow!80!white}63 & \cellcolor{mediumgreen!67!mediumyellow!80!white}86 & \cellcolor{mediumgreen!87!mediumyellow!80!white}94 & \cellcolor{mediumgreen!97!mediumyellow!80!white}98 & \cellcolor{mediumgreen!100!mediumyellow!80!white}99 & \cellcolor{mediumgreen!7!mediumyellow!80!white}63 & \cellcolor{mediumgreen!68!mediumyellow!80!white}87 & \cellcolor{mediumgreen!87!mediumyellow!80!white}94 & \cellcolor{mediumgreen!98!mediumyellow!80!white}99 & \cellcolor{mediumgreen!100!mediumyellow!80!white}99 \\
 1K-L-R & \cellcolor{mediumyellow!75!mediumred!80!white}50 & \cellcolor{mediumgreen!51!mediumyellow!80!white}80 & \cellcolor{mediumgreen!81!mediumyellow!80!white}92 & \cellcolor{mediumgreen!98!mediumyellow!80!white}99 & \cellcolor{mediumgreen!100!mediumyellow!80!white}99 & \cellcolor{mediumgreen!10!mediumyellow!80!white}64 & \cellcolor{mediumgreen!72!mediumyellow!80!white}88 & \cellcolor{mediumgreen!91!mediumyellow!80!white}96 & \cellcolor{mediumgreen!99!mediumyellow!80!white}99 & \cellcolor{mediumgreen!100!mediumyellow!80!white}99 & \cellcolor{mediumgreen!14!mediumyellow!80!white}65 & \cellcolor{mediumgreen!77!mediumyellow!80!white}90 & \cellcolor{mediumgreen!93!mediumyellow!80!white}97 & \cellcolor{mediumgreen!99!mediumyellow!80!white}99 & \cellcolor{mediumgreen!100!mediumyellow!80!white}99 & \cellcolor{mediumgreen!19!mediumyellow!80!white}67 & \cellcolor{mediumgreen!71!mediumyellow!80!white}88 & \cellcolor{mediumgreen!85!mediumyellow!80!white}94 & \cellcolor{mediumgreen!98!mediumyellow!80!white}99 & \cellcolor{mediumgreen!100!mediumyellow!80!white}99 & \cellcolor{mediumgreen!6!mediumyellow!80!white}62 & \cellcolor{mediumgreen!66!mediumyellow!80!white}86 & \cellcolor{mediumgreen!85!mediumyellow!80!white}93 & \cellcolor{mediumgreen!97!mediumyellow!80!white}98 & \cellcolor{mediumgreen!100!mediumyellow!80!white}99 & \cellcolor{mediumgreen!6!mediumyellow!80!white}62 & \cellcolor{mediumgreen!61!mediumyellow!80!white}84 & \cellcolor{mediumgreen!85!mediumyellow!80!white}94 & \cellcolor{mediumgreen!97!mediumyellow!80!white}98 & \cellcolor{mediumgreen!100!mediumyellow!80!white}99 \\
 10K-S-N & \cellcolor{mediumyellow!2!mediumred!80!white}21 & \cellcolor{mediumyellow!46!mediumred!80!white}39 & \cellcolor{mediumyellow!87!mediumred!80!white}55 & \cellcolor{mediumgreen!20!mediumyellow!80!white}68 & \cellcolor{mediumgreen!46!mediumyellow!80!white}78 & \cellcolor{mediumyellow!15!mediumred!80!white}27 & \cellcolor{mediumyellow!59!mediumred!80!white}44 & \cellcolor{mediumyellow!98!mediumred!80!white}59 & \cellcolor{mediumgreen!30!mediumyellow!80!white}72 & \cellcolor{mediumgreen!53!mediumyellow!80!white}81 & \cellcolor{mediumyellow!19!mediumred!80!white}28 & \cellcolor{mediumyellow!63!mediumred!80!white}46 & \cellcolor{mediumgreen!2!mediumyellow!80!white}61 & \cellcolor{mediumgreen!34!mediumyellow!80!white}73 & \cellcolor{mediumgreen!51!mediumyellow!80!white}80 & \cellcolor{mediumyellow!22!mediumred!80!white}29 & \cellcolor{mediumyellow!66!mediumred!80!white}47 & \cellcolor{mediumyellow!98!mediumred!80!white}59 & \cellcolor{mediumgreen!22!mediumyellow!80!white}69 & \cellcolor{mediumgreen!43!mediumyellow!80!white}77 & \cellcolor{mediumyellow!24!mediumred!80!white}30 & \cellcolor{mediumyellow!66!mediumred!80!white}47 & \cellcolor{mediumyellow!93!mediumred!80!white}57 & \cellcolor{mediumgreen!20!mediumyellow!80!white}68 & \cellcolor{mediumgreen!42!mediumyellow!80!white}77 & \cellcolor{mediumyellow!26!mediumred!80!white}31 & \cellcolor{mediumyellow!61!mediumred!80!white}45 & \cellcolor{mediumyellow!94!mediumred!80!white}58 & \cellcolor{mediumgreen!21!mediumyellow!80!white}68 & \cellcolor{mediumgreen!40!mediumyellow!80!white}76 \\
 10K-S-R & \cellcolor{mediumyellow!0!mediumred!80!white}20 & \cellcolor{mediumyellow!49!mediumred!80!white}40 & \cellcolor{mediumyellow!91!mediumred!80!white}56 & \cellcolor{mediumgreen!26!mediumyellow!80!white}70 & \cellcolor{mediumgreen!50!mediumyellow!80!white}80 & \cellcolor{mediumyellow!16!mediumred!80!white}27 & \cellcolor{mediumyellow!63!mediumred!80!white}45 & \cellcolor{mediumgreen!3!mediumyellow!80!white}61 & \cellcolor{mediumgreen!35!mediumyellow!80!white}74 & \cellcolor{mediumgreen!56!mediumyellow!80!white}82 & \cellcolor{mediumyellow!22!mediumred!80!white}29 & \cellcolor{mediumyellow!67!mediumred!80!white}47 & \cellcolor{mediumgreen!6!mediumyellow!80!white}62 & \cellcolor{mediumgreen!35!mediumyellow!80!white}74 & \cellcolor{mediumgreen!56!mediumyellow!80!white}82 & \cellcolor{mediumyellow!24!mediumred!80!white}30 & \cellcolor{mediumyellow!70!mediumred!80!white}48 & \cellcolor{mediumgreen!2!mediumyellow!80!white}61 & \cellcolor{mediumgreen!26!mediumyellow!80!white}70 & \cellcolor{mediumgreen!46!mediumyellow!80!white}78 & \cellcolor{mediumyellow!26!mediumred!80!white}31 & \cellcolor{mediumyellow!69!mediumred!80!white}48 & \cellcolor{mediumyellow!96!mediumred!80!white}58 & \cellcolor{mediumgreen!19!mediumyellow!80!white}67 & \cellcolor{mediumgreen!40!mediumyellow!80!white}76 & \cellcolor{mediumyellow!27!mediumred!80!white}31 & \cellcolor{mediumyellow!64!mediumred!80!white}46 & \cellcolor{mediumyellow!92!mediumred!80!white}57 & \cellcolor{mediumgreen!16!mediumyellow!80!white}66 & \cellcolor{mediumgreen!36!mediumyellow!80!white}74 \\
 10K-M-N & \cellcolor{mediumyellow!5!mediumred!80!white}23 & \cellcolor{mediumyellow!53!mediumred!80!white}41 & \cellcolor{mediumyellow!94!mediumred!80!white}58 & \cellcolor{mediumgreen!28!mediumyellow!80!white}71 & \cellcolor{mediumgreen!53!mediumyellow!80!white}81 & \cellcolor{mediumyellow!19!mediumred!80!white}28 & \cellcolor{mediumyellow!65!mediumred!80!white}46 & \cellcolor{mediumgreen!4!mediumyellow!80!white}61 & \cellcolor{mediumgreen!38!mediumyellow!80!white}75 & \cellcolor{mediumgreen!62!mediumyellow!80!white}84 & \cellcolor{mediumyellow!23!mediumred!80!white}29 & \cellcolor{mediumyellow!69!mediumred!80!white}48 & \cellcolor{mediumgreen!10!mediumyellow!80!white}64 & \cellcolor{mediumgreen!41!mediumyellow!80!white}76 & \cellcolor{mediumgreen!58!mediumyellow!80!white}83 & \cellcolor{mediumyellow!25!mediumred!80!white}30 & \cellcolor{mediumyellow!72!mediumred!80!white}49 & \cellcolor{mediumgreen!3!mediumyellow!80!white}61 & \cellcolor{mediumgreen!29!mediumyellow!80!white}71 & \cellcolor{mediumgreen!51!mediumyellow!80!white}80 & \cellcolor{mediumyellow!28!mediumred!80!white}31 & \cellcolor{mediumyellow!70!mediumred!80!white}48 & \cellcolor{mediumyellow!99!mediumred!80!white}60 & \cellcolor{mediumgreen!27!mediumyellow!80!white}71 & \cellcolor{mediumgreen!50!mediumyellow!80!white}80 & \cellcolor{mediumyellow!29!mediumred!80!white}32 & \cellcolor{mediumyellow!66!mediumred!80!white}47 & \cellcolor{mediumgreen!1!mediumyellow!80!white}60 & \cellcolor{mediumgreen!28!mediumyellow!80!white}71 & \cellcolor{mediumgreen!48!mediumyellow!80!white}79 \\
 10K-M-R & \cellcolor{mediumyellow!1!mediumred!80!white}21 & \cellcolor{mediumyellow!52!mediumred!80!white}41 & \cellcolor{mediumyellow!98!mediumred!80!white}59 & \cellcolor{mediumgreen!32!mediumyellow!80!white}73 & \cellcolor{mediumgreen!58!mediumyellow!80!white}83 & \cellcolor{mediumyellow!19!mediumred!80!white}28 & \cellcolor{mediumyellow!68!mediumred!80!white}48 & \cellcolor{mediumgreen!7!mediumyellow!80!white}63 & \cellcolor{mediumgreen!41!mediumyellow!80!white}76 & \cellcolor{mediumgreen!64!mediumyellow!80!white}85 & \cellcolor{mediumyellow!24!mediumred!80!white}30 & \cellcolor{mediumyellow!72!mediumred!80!white}49 & \cellcolor{mediumgreen!13!mediumyellow!80!white}65 & \cellcolor{mediumgreen!44!mediumyellow!80!white}77 & \cellcolor{mediumgreen!62!mediumyellow!80!white}84 & \cellcolor{mediumyellow!26!mediumred!80!white}31 & \cellcolor{mediumyellow!75!mediumred!80!white}50 & \cellcolor{mediumgreen!6!mediumyellow!80!white}63 & \cellcolor{mediumgreen!32!mediumyellow!80!white}72 & \cellcolor{mediumgreen!52!mediumyellow!80!white}81 & \cellcolor{mediumyellow!29!mediumred!80!white}32 & \cellcolor{mediumyellow!73!mediumred!80!white}49 & \cellcolor{mediumgreen!0!mediumyellow!80!white}60 & \cellcolor{mediumgreen!25!mediumyellow!80!white}70 & \cellcolor{mediumgreen!47!mediumyellow!80!white}79 & \cellcolor{mediumyellow!30!mediumred!80!white}32 & \cellcolor{mediumyellow!68!mediumred!80!white}47 & \cellcolor{mediumyellow!97!mediumred!80!white}59 & \cellcolor{mediumgreen!22!mediumyellow!80!white}69 & \cellcolor{mediumgreen!43!mediumyellow!80!white}77 \\
 10K-L-N & \cellcolor{mediumyellow!39!mediumred!80!white}36 & \cellcolor{mediumgreen!8!mediumyellow!80!white}63 & \cellcolor{mediumgreen!61!mediumyellow!80!white}84 & \cellcolor{mediumgreen!86!mediumyellow!80!white}94 & \cellcolor{mediumgreen!97!mediumyellow!80!white}98 & \cellcolor{mediumyellow!59!mediumred!80!white}44 & \cellcolor{mediumgreen!27!mediumyellow!80!white}71 & \cellcolor{mediumgreen!73!mediumyellow!80!white}89 & \cellcolor{mediumgreen!92!mediumyellow!80!white}96 & \cellcolor{mediumgreen!98!mediumyellow!80!white}99 & \cellcolor{mediumyellow!64!mediumred!80!white}46 & \cellcolor{mediumgreen!32!mediumyellow!80!white}73 & \cellcolor{mediumgreen!70!mediumyellow!80!white}87 & \cellcolor{mediumgreen!87!mediumyellow!80!white}94 & \cellcolor{mediumgreen!97!mediumyellow!80!white}98 & \cellcolor{mediumyellow!67!mediumred!80!white}47 & \cellcolor{mediumgreen!23!mediumyellow!80!white}69 & \cellcolor{mediumgreen!60!mediumyellow!80!white}83 & \cellcolor{mediumgreen!82!mediumyellow!80!white}92 & \cellcolor{mediumgreen!95!mediumyellow!80!white}97 & \cellcolor{mediumyellow!65!mediumred!80!white}46 & \cellcolor{mediumgreen!19!mediumyellow!80!white}67 & \cellcolor{mediumgreen!60!mediumyellow!80!white}84 & \cellcolor{mediumgreen!79!mediumyellow!80!white}91 & \cellcolor{mediumgreen!94!mediumyellow!80!white}97 & \cellcolor{mediumyellow!61!mediumred!80!white}45 & \cellcolor{mediumgreen!19!mediumyellow!80!white}68 & \cellcolor{mediumgreen!58!mediumyellow!80!white}83 & \cellcolor{mediumgreen!79!mediumyellow!80!white}91 & \cellcolor{mediumgreen!93!mediumyellow!80!white}97 \\
 10K-L-R & \cellcolor{mediumyellow!24!mediumred!80!white}30 & \cellcolor{mediumyellow!81!mediumred!80!white}52 & \cellcolor{mediumgreen!40!mediumyellow!80!white}76 & \cellcolor{mediumgreen!72!mediumyellow!80!white}89 & \cellcolor{mediumgreen!91!mediumyellow!80!white}96 & \cellcolor{mediumyellow!43!mediumred!80!white}37 & \cellcolor{mediumgreen!5!mediumyellow!80!white}62 & \cellcolor{mediumgreen!55!mediumyellow!80!white}81 & \cellcolor{mediumgreen!82!mediumyellow!80!white}92 & \cellcolor{mediumgreen!94!mediumyellow!80!white}97 & \cellcolor{mediumyellow!48!mediumred!80!white}40 & \cellcolor{mediumgreen!11!mediumyellow!80!white}64 & \cellcolor{mediumgreen!57!mediumyellow!80!white}83 & \cellcolor{mediumgreen!80!mediumyellow!80!white}91 & \cellcolor{mediumgreen!92!mediumyellow!80!white}96 & \cellcolor{mediumyellow!52!mediumred!80!white}41 & \cellcolor{mediumgreen!10!mediumyellow!80!white}64 & \cellcolor{mediumgreen!41!mediumyellow!80!white}76 & \cellcolor{mediumgreen!68!mediumyellow!80!white}87 & \cellcolor{mediumgreen!87!mediumyellow!80!white}94 & \cellcolor{mediumyellow!54!mediumred!80!white}42 & \cellcolor{mediumgreen!2!mediumyellow!80!white}61 & \cellcolor{mediumgreen!35!mediumyellow!80!white}74 & \cellcolor{mediumgreen!61!mediumyellow!80!white}84 & \cellcolor{mediumgreen!84!mediumyellow!80!white}93 & \cellcolor{mediumyellow!53!mediumred!80!white}41 & \cellcolor{mediumyellow!97!mediumred!80!white}59 & \cellcolor{mediumgreen!33!mediumyellow!80!white}73 & \cellcolor{mediumgreen!60!mediumyellow!80!white}84 & \cellcolor{mediumgreen!82!mediumyellow!80!white}92 \\
\end{tabular}
}
\end{table}

Table~\ref{tab:mu_ITRSallscenarios} shows the results, with the entries being the unit throughput rate as a percentage of the upper bound.
In each cell the result of the best performing RC for the respective scenario and station / robot configuration is shown.
The unit throughput rate scales well when adding more pick stations, the scaling is (almost) completely independent of the scenario characteristics.
However, the necessary number of robots to achieve a given unit throughput rate greatly depends on the scenario characteristics, e.g., for more SKUs more robots are necessary to achieve a high unit throughput rate.
The number of SKUs, does have a major impact on performance overall, where the main reason is that pile-on is considerably lower for the 10000 SKU scenarios.
A reason for this is the lower likeliness to have a pod with a good combination of SKUs matching the orders of the pick stations available.
Thereby, if larger orders have to be processed with the system, this helps mitigating the negative effect of handling lots of SKUs.
The main reason for this are the larger number of order lines active at a station when picking larger orders.
I.e., more open order lines increase the likeliness of having a well matching pod available for the inventory required at a pick station.
Processing return orders has an increased negative effect, if the order size of customer orders is large.
However, in general, whether return orders are processed has a lesser effect on throughput performance than the other warehouse scenario variations.
The reason behind this may be that even though approximately 19.76 \% more time is spent on replenishment operations by the robots when compared to the scenarios without return order processing, replenishment operations are overall quick enough to mitigate the effect.
Replenishment operations only consume 20.29 \% out of the overall time consumed by the robots in average across all phase 2 simulation runs.
Furthermore, we can conclude that with 1000 SKUs, the unit throughput rates are close to their theoretical maximum even with relatively few robots per stations.\\
Table~\ref{tab:mu_phase2_results_RC} shows the unit throughput rate score for the RCs for all combinations of number of robots ($n_r$) and number of stations ($n_s$), averaged across WSs and presented as whole percentages.
From Table~\ref{tab:mu_phase2_results_RC} we can see that the Ranked RCs from phase 1 perform similarly and better than the benchmark RCs.
Among the benchmark RCs, the Greedy benchmark outperforms the others consistently across all settings and is the only one whose unit throughput rate scores approached those of the ranked RCs.

\begin{table}[htbp]
	\begin{center}
		\caption{Unit throughput rate scores for the RCs in phase 2 (green $\equiv$ best, red $\equiv$ worst)}
		\label{tab:mu_phase2_results_RC}
		\setlength{\tabcolsep}{1mm}
			\begin{tabular}{r@{ , }l|cccc|cccccc}
			\multicolumn{2}{c|}{} & \multicolumn{4}{c|}{Ranked RCs} & \multicolumn{6}{c}{Benchmark RCs}\\
			 ($n_s$ & $n_r$) & 1 & 2 & 3 & 4 & Demand & Speed & Nearest & Class & Greedy & Random \\
			\hline
			(1 & 2) & \cellcolor{mediumyellow!47!mediumred!80!white}30 & \cellcolor{mediumyellow!42!mediumred!80!white}28 & \cellcolor{mediumyellow!52!mediumred!80!white}32 & \cellcolor{mediumyellow!42!mediumred!80!white}28 & \cellcolor{mediumyellow!7!mediumred!80!white}14 & \cellcolor{mediumyellow!30!mediumred!80!white}23 & \cellcolor{mediumyellow!17!mediumred!80!white}18 & \cellcolor{mediumyellow!35!mediumred!80!white}25 & \cellcolor{mediumyellow!57!mediumred!80!white}34 & \cellcolor{mediumyellow!0!mediumred!80!white}11 \\
			(1 & 3) & \cellcolor{mediumgreen!21!mediumyellow!80!white}60 & \cellcolor{mediumgreen!19!mediumyellow!80!white}59 & \cellcolor{mediumgreen!23!mediumyellow!80!white}61 & \cellcolor{mediumgreen!16!mediumyellow!80!white}58 & \cellcolor{mediumyellow!32!mediumred!80!white}24 & \cellcolor{mediumyellow!72!mediumred!80!white}40 & \cellcolor{mediumyellow!57!mediumred!80!white}34 & \cellcolor{mediumyellow!77!mediumred!80!white}42 & \cellcolor{mediumgreen!14!mediumyellow!80!white}57 & \cellcolor{mediumyellow!30!mediumred!80!white}23 \\
			(1 & 4) & \cellcolor{mediumgreen!60!mediumyellow!80!white}76 & \cellcolor{mediumgreen!58!mediumyellow!80!white}75 & \cellcolor{mediumgreen!60!mediumyellow!80!white}76 & \cellcolor{mediumgreen!56!mediumyellow!80!white}74 & \cellcolor{mediumyellow!64!mediumred!80!white}37 & \cellcolor{mediumgreen!14!mediumyellow!80!white}57 & \cellcolor{mediumyellow!94!mediumred!80!white}49 & \cellcolor{mediumgreen!9!mediumyellow!80!white}55 & \cellcolor{mediumgreen!51!mediumyellow!80!white}72 & \cellcolor{mediumyellow!30!mediumred!80!white}23 \\
			(1 & 5) & \cellcolor{mediumgreen!85!mediumyellow!80!white}86 & \cellcolor{mediumgreen!85!mediumyellow!80!white}86 & \cellcolor{mediumgreen!85!mediumyellow!80!white}86 & \cellcolor{mediumgreen!80!mediumyellow!80!white}84 & \cellcolor{mediumyellow!89!mediumred!80!white}47 & \cellcolor{mediumgreen!46!mediumyellow!80!white}70 & \cellcolor{mediumgreen!26!mediumyellow!80!white}62 & \cellcolor{mediumgreen!36!mediumyellow!80!white}66 & \cellcolor{mediumgreen!75!mediumyellow!80!white}82 & \cellcolor{mediumyellow!57!mediumred!80!white}34 \\
			(1 & 6) & \cellcolor{mediumgreen!98!mediumyellow!80!white}91 & \cellcolor{mediumgreen!98!mediumyellow!80!white}91 & \cellcolor{mediumgreen!98!mediumyellow!80!white}91 & \cellcolor{mediumgreen!95!mediumyellow!80!white}90 & \cellcolor{mediumgreen!16!mediumyellow!80!white}58 & \cellcolor{mediumgreen!68!mediumyellow!80!white}79 & \cellcolor{mediumgreen!53!mediumyellow!80!white}73 & \cellcolor{mediumgreen!56!mediumyellow!80!white}74 & \cellcolor{mediumgreen!88!mediumyellow!80!white}87 & \cellcolor{mediumyellow!84!mediumred!80!white}45 \\
			(2 & 2) & \cellcolor{mediumyellow!77!mediumred!80!white}42 & \cellcolor{mediumyellow!74!mediumred!80!white}41 & \cellcolor{mediumyellow!79!mediumred!80!white}43 & \cellcolor{mediumyellow!69!mediumred!80!white}39 & \cellcolor{mediumyellow!17!mediumred!80!white}18 & \cellcolor{mediumyellow!42!mediumred!80!white}28 & \cellcolor{mediumyellow!35!mediumred!80!white}25 & \cellcolor{mediumyellow!52!mediumred!80!white}32 & \cellcolor{mediumyellow!77!mediumred!80!white}42 & \cellcolor{mediumyellow!0!mediumred!80!white}11 \\
			(2 & 3) & \cellcolor{mediumgreen!41!mediumyellow!80!white}68 & \cellcolor{mediumgreen!38!mediumyellow!80!white}67 & \cellcolor{mediumgreen!41!mediumyellow!80!white}68 & \cellcolor{mediumgreen!33!mediumyellow!80!white}65 & \cellcolor{mediumyellow!44!mediumred!80!white}29 & \cellcolor{mediumyellow!89!mediumred!80!white}47 & \cellcolor{mediumyellow!72!mediumred!80!white}40 & \cellcolor{mediumyellow!89!mediumred!80!white}47 & \cellcolor{mediumgreen!31!mediumyellow!80!white}64 & \cellcolor{mediumyellow!30!mediumred!80!white}23 \\
			(2 & 4) & \cellcolor{mediumgreen!73!mediumyellow!80!white}81 & \cellcolor{mediumgreen!70!mediumyellow!80!white}80 & \cellcolor{mediumgreen!70!mediumyellow!80!white}80 & \cellcolor{mediumgreen!65!mediumyellow!80!white}78 & \cellcolor{mediumyellow!72!mediumred!80!white}40 & \cellcolor{mediumgreen!28!mediumyellow!80!white}63 & \cellcolor{mediumgreen!9!mediumyellow!80!white}55 & \cellcolor{mediumgreen!19!mediumyellow!80!white}59 & \cellcolor{mediumgreen!60!mediumyellow!80!white}76 & \cellcolor{mediumyellow!44!mediumred!80!white}29 \\
			(2 & 5) & \cellcolor{mediumgreen!90!mediumyellow!80!white}88 & \cellcolor{mediumgreen!90!mediumyellow!80!white}88 & \cellcolor{mediumgreen!90!mediumyellow!80!white}88 & \cellcolor{mediumgreen!85!mediumyellow!80!white}86 & \cellcolor{mediumyellow!99!mediumred!80!white}51 & \cellcolor{mediumgreen!56!mediumyellow!80!white}74 & \cellcolor{mediumgreen!38!mediumyellow!80!white}67 & \cellcolor{mediumgreen!43!mediumyellow!80!white}69 & \cellcolor{mediumgreen!80!mediumyellow!80!white}84 & \cellcolor{mediumyellow!64!mediumred!80!white}37 \\
			(2 & 6) & \cellcolor{mediumgreen!100!mediumyellow!80!white}92 & \cellcolor{mediumgreen!100!mediumyellow!80!white}92 & \cellcolor{mediumgreen!100!mediumyellow!80!white}92 & \cellcolor{mediumgreen!98!mediumyellow!80!white}91 & \cellcolor{mediumgreen!26!mediumyellow!80!white}62 & \cellcolor{mediumgreen!78!mediumyellow!80!white}83 & \cellcolor{mediumgreen!63!mediumyellow!80!white}77 & \cellcolor{mediumgreen!60!mediumyellow!80!white}76 & \cellcolor{mediumgreen!93!mediumyellow!80!white}89 & \cellcolor{mediumyellow!89!mediumred!80!white}47 \\
			(3 & 2) & \cellcolor{mediumyellow!84!mediumred!80!white}45 & \cellcolor{mediumyellow!79!mediumred!80!white}43 & \cellcolor{mediumyellow!86!mediumred!80!white}46 & \cellcolor{mediumyellow!77!mediumred!80!white}42 & \cellcolor{mediumyellow!20!mediumred!80!white}19 & \cellcolor{mediumyellow!52!mediumred!80!white}32 & \cellcolor{mediumyellow!40!mediumred!80!white}27 & \cellcolor{mediumyellow!52!mediumred!80!white}32 & \cellcolor{mediumyellow!81!mediumred!80!white}44 & \cellcolor{mediumyellow!10!mediumred!80!white}15 \\
			(3 & 3) & \cellcolor{mediumgreen!46!mediumyellow!80!white}70 & \cellcolor{mediumgreen!43!mediumyellow!80!white}69 & \cellcolor{mediumgreen!46!mediumyellow!80!white}70 & \cellcolor{mediumgreen!36!mediumyellow!80!white}66 & \cellcolor{mediumyellow!47!mediumred!80!white}30 & \cellcolor{mediumyellow!96!mediumred!80!white}50 & \cellcolor{mediumyellow!79!mediumred!80!white}43 & \cellcolor{mediumyellow!91!mediumred!80!white}48 & \cellcolor{mediumgreen!33!mediumyellow!80!white}65 & \cellcolor{mediumyellow!30!mediumred!80!white}23 \\
			(3 & 4) & \cellcolor{mediumgreen!73!mediumyellow!80!white}81 & \cellcolor{mediumgreen!73!mediumyellow!80!white}81 & \cellcolor{mediumgreen!73!mediumyellow!80!white}81 & \cellcolor{mediumgreen!68!mediumyellow!80!white}79 & \cellcolor{mediumyellow!77!mediumred!80!white}42 & \cellcolor{mediumgreen!33!mediumyellow!80!white}65 & \cellcolor{mediumgreen!14!mediumyellow!80!white}57 & \cellcolor{mediumgreen!21!mediumyellow!80!white}60 & \cellcolor{mediumgreen!63!mediumyellow!80!white}77 & \cellcolor{mediumyellow!49!mediumred!80!white}31 \\
			(3 & 5) & \cellcolor{mediumgreen!90!mediumyellow!80!white}88 & \cellcolor{mediumgreen!90!mediumyellow!80!white}88 & \cellcolor{mediumgreen!90!mediumyellow!80!white}88 & \cellcolor{mediumgreen!85!mediumyellow!80!white}86 & \cellcolor{mediumgreen!1!mediumyellow!80!white}52 & \cellcolor{mediumgreen!58!mediumyellow!80!white}75 & \cellcolor{mediumgreen!41!mediumyellow!80!white}68 & \cellcolor{mediumgreen!43!mediumyellow!80!white}69 & \cellcolor{mediumgreen!80!mediumyellow!80!white}84 & \cellcolor{mediumyellow!74!mediumred!80!white}41 \\
			(3 & 6) & \cellcolor{mediumgreen!100!mediumyellow!80!white}92 & \cellcolor{mediumgreen!100!mediumyellow!80!white}92 & \cellcolor{mediumgreen!100!mediumyellow!80!white}92 & \cellcolor{mediumgreen!95!mediumyellow!80!white}90 & \cellcolor{mediumgreen!26!mediumyellow!80!white}62 & \cellcolor{mediumgreen!78!mediumyellow!80!white}83 & \cellcolor{mediumgreen!63!mediumyellow!80!white}77 & \cellcolor{mediumgreen!60!mediumyellow!80!white}76 & \cellcolor{mediumgreen!93!mediumyellow!80!white}89 & \cellcolor{mediumyellow!96!mediumred!80!white}50 \\
			(4 & 2) & \cellcolor{mediumyellow!84!mediumred!80!white}45 & \cellcolor{mediumyellow!81!mediumred!80!white}44 & \cellcolor{mediumyellow!89!mediumred!80!white}47 & \cellcolor{mediumyellow!77!mediumred!80!white}42 & \cellcolor{mediumyellow!22!mediumred!80!white}20 & \cellcolor{mediumyellow!57!mediumred!80!white}34 & \cellcolor{mediumyellow!44!mediumred!80!white}29 & \cellcolor{mediumyellow!57!mediumred!80!white}34 & \cellcolor{mediumyellow!84!mediumred!80!white}45 & \cellcolor{mediumyellow!7!mediumred!80!white}14 \\
			(4 & 3) & \cellcolor{mediumgreen!41!mediumyellow!80!white}68 & \cellcolor{mediumgreen!38!mediumyellow!80!white}67 & \cellcolor{mediumgreen!43!mediumyellow!80!white}69 & \cellcolor{mediumgreen!33!mediumyellow!80!white}65 & \cellcolor{mediumyellow!49!mediumred!80!white}31 & \cellcolor{mediumyellow!99!mediumred!80!white}51 & \cellcolor{mediumyellow!81!mediumred!80!white}44 & \cellcolor{mediumyellow!94!mediumred!80!white}49 & \cellcolor{mediumgreen!31!mediumyellow!80!white}64 & \cellcolor{mediumyellow!30!mediumred!80!white}23 \\
			(4 & 4) & \cellcolor{mediumgreen!65!mediumyellow!80!white}78 & \cellcolor{mediumgreen!65!mediumyellow!80!white}78 & \cellcolor{mediumgreen!65!mediumyellow!80!white}78 & \cellcolor{mediumgreen!60!mediumyellow!80!white}76 & \cellcolor{mediumyellow!77!mediumred!80!white}42 & \cellcolor{mediumgreen!31!mediumyellow!80!white}64 & \cellcolor{mediumgreen!11!mediumyellow!80!white}56 & \cellcolor{mediumgreen!19!mediumyellow!80!white}59 & \cellcolor{mediumgreen!56!mediumyellow!80!white}74 & \cellcolor{mediumyellow!57!mediumred!80!white}34 \\
			(4 & 5) & \cellcolor{mediumgreen!85!mediumyellow!80!white}86 & \cellcolor{mediumgreen!85!mediumyellow!80!white}86 & \cellcolor{mediumgreen!85!mediumyellow!80!white}86 & \cellcolor{mediumgreen!80!mediumyellow!80!white}84 & \cellcolor{mediumyellow!99!mediumred!80!white}51 & \cellcolor{mediumgreen!56!mediumyellow!80!white}74 & \cellcolor{mediumgreen!36!mediumyellow!80!white}66 & \cellcolor{mediumgreen!41!mediumyellow!80!white}68 & \cellcolor{mediumgreen!75!mediumyellow!80!white}82 & \cellcolor{mediumyellow!79!mediumred!80!white}43 \\
			(4 & 6) & \cellcolor{mediumgreen!98!mediumyellow!80!white}91 & \cellcolor{mediumgreen!95!mediumyellow!80!white}90 & \cellcolor{mediumgreen!95!mediumyellow!80!white}90 & \cellcolor{mediumgreen!93!mediumyellow!80!white}89 & \cellcolor{mediumgreen!21!mediumyellow!80!white}60 & \cellcolor{mediumgreen!73!mediumyellow!80!white}81 & \cellcolor{mediumgreen!56!mediumyellow!80!white}74 & \cellcolor{mediumgreen!56!mediumyellow!80!white}74 & \cellcolor{mediumgreen!88!mediumyellow!80!white}87 & \cellcolor{mediumgreen!4!mediumyellow!80!white}53 \\
			(5 & 2) & \cellcolor{mediumyellow!79!mediumred!80!white}43 & \cellcolor{mediumyellow!77!mediumred!80!white}42 & \cellcolor{mediumyellow!84!mediumred!80!white}45 & \cellcolor{mediumyellow!72!mediumred!80!white}40 & \cellcolor{mediumyellow!22!mediumred!80!white}20 & \cellcolor{mediumyellow!57!mediumred!80!white}34 & \cellcolor{mediumyellow!44!mediumred!80!white}29 & \cellcolor{mediumyellow!59!mediumred!80!white}35 & \cellcolor{mediumyellow!79!mediumred!80!white}43 & \cellcolor{mediumyellow!7!mediumred!80!white}14 \\
			(5 & 3) & \cellcolor{mediumgreen!38!mediumyellow!80!white}67 & \cellcolor{mediumgreen!36!mediumyellow!80!white}66 & \cellcolor{mediumgreen!41!mediumyellow!80!white}68 & \cellcolor{mediumgreen!31!mediumyellow!80!white}64 & \cellcolor{mediumyellow!49!mediumred!80!white}31 & \cellcolor{mediumyellow!99!mediumred!80!white}51 & \cellcolor{mediumyellow!81!mediumred!80!white}44 & \cellcolor{mediumyellow!91!mediumred!80!white}48 & \cellcolor{mediumgreen!28!mediumyellow!80!white}63 & \cellcolor{mediumyellow!32!mediumred!80!white}24 \\
			(5 & 4) & \cellcolor{mediumgreen!65!mediumyellow!80!white}78 & \cellcolor{mediumgreen!60!mediumyellow!80!white}76 & \cellcolor{mediumgreen!63!mediumyellow!80!white}77 & \cellcolor{mediumgreen!58!mediumyellow!80!white}75 & \cellcolor{mediumyellow!74!mediumred!80!white}41 & \cellcolor{mediumgreen!28!mediumyellow!80!white}63 & \cellcolor{mediumgreen!9!mediumyellow!80!white}55 & \cellcolor{mediumgreen!16!mediumyellow!80!white}58 & \cellcolor{mediumgreen!53!mediumyellow!80!white}73 & \cellcolor{mediumyellow!59!mediumred!80!white}35 \\
			(5 & 5) & \cellcolor{mediumgreen!83!mediumyellow!80!white}85 & \cellcolor{mediumgreen!80!mediumyellow!80!white}84 & \cellcolor{mediumgreen!83!mediumyellow!80!white}85 & \cellcolor{mediumgreen!78!mediumyellow!80!white}83 & \cellcolor{mediumyellow!94!mediumred!80!white}49 & \cellcolor{mediumgreen!51!mediumyellow!80!white}72 & \cellcolor{mediumgreen!28!mediumyellow!80!white}63 & \cellcolor{mediumgreen!36!mediumyellow!80!white}66 & \cellcolor{mediumgreen!73!mediumyellow!80!white}81 & \cellcolor{mediumyellow!84!mediumred!80!white}45 \\
			(5 & 6) & \cellcolor{mediumgreen!95!mediumyellow!80!white}90 & \cellcolor{mediumgreen!95!mediumyellow!80!white}90 & \cellcolor{mediumgreen!95!mediumyellow!80!white}90 & \cellcolor{mediumgreen!90!mediumyellow!80!white}88 & \cellcolor{mediumgreen!16!mediumyellow!80!white}58 & \cellcolor{mediumgreen!70!mediumyellow!80!white}80 & \cellcolor{mediumgreen!51!mediumyellow!80!white}72 & \cellcolor{mediumgreen!53!mediumyellow!80!white}73 & \cellcolor{mediumgreen!85!mediumyellow!80!white}86 & \cellcolor{mediumgreen!6!mediumyellow!80!white}54 \\
			(6 & 2) & \cellcolor{mediumyellow!81!mediumred!80!white}44 & \cellcolor{mediumyellow!77!mediumred!80!white}42 & \cellcolor{mediumyellow!84!mediumred!80!white}45 & \cellcolor{mediumyellow!74!mediumred!80!white}41 & \cellcolor{mediumyellow!22!mediumred!80!white}20 & \cellcolor{mediumyellow!59!mediumred!80!white}35 & \cellcolor{mediumyellow!47!mediumred!80!white}30 & \cellcolor{mediumyellow!59!mediumred!80!white}35 & \cellcolor{mediumyellow!79!mediumred!80!white}43 & \cellcolor{mediumyellow!10!mediumred!80!white}15 \\
			(6 & 3) & \cellcolor{mediumgreen!36!mediumyellow!80!white}66 & \cellcolor{mediumgreen!31!mediumyellow!80!white}64 & \cellcolor{mediumgreen!36!mediumyellow!80!white}66 & \cellcolor{mediumgreen!28!mediumyellow!80!white}63 & \cellcolor{mediumyellow!49!mediumred!80!white}31 & \cellcolor{mediumyellow!99!mediumred!80!white}51 & \cellcolor{mediumyellow!79!mediumred!80!white}43 & \cellcolor{mediumyellow!91!mediumred!80!white}48 & \cellcolor{mediumgreen!26!mediumyellow!80!white}62 & \cellcolor{mediumyellow!35!mediumred!80!white}25 \\
			(6 & 4) & \cellcolor{mediumgreen!63!mediumyellow!80!white}77 & \cellcolor{mediumgreen!60!mediumyellow!80!white}76 & \cellcolor{mediumgreen!63!mediumyellow!80!white}77 & \cellcolor{mediumgreen!58!mediumyellow!80!white}75 & \cellcolor{mediumyellow!72!mediumred!80!white}40 & \cellcolor{mediumgreen!26!mediumyellow!80!white}62 & \cellcolor{mediumgreen!4!mediumyellow!80!white}53 & \cellcolor{mediumgreen!16!mediumyellow!80!white}58 & \cellcolor{mediumgreen!53!mediumyellow!80!white}73 & \cellcolor{mediumyellow!59!mediumred!80!white}35 \\
			(6 & 5) & \cellcolor{mediumgreen!83!mediumyellow!80!white}85 & \cellcolor{mediumgreen!80!mediumyellow!80!white}84 & \cellcolor{mediumgreen!83!mediumyellow!80!white}85 & \cellcolor{mediumgreen!78!mediumyellow!80!white}83 & \cellcolor{mediumyellow!91!mediumred!80!white}48 & \cellcolor{mediumgreen!48!mediumyellow!80!white}71 & \cellcolor{mediumgreen!26!mediumyellow!80!white}62 & \cellcolor{mediumgreen!33!mediumyellow!80!white}65 & \cellcolor{mediumgreen!70!mediumyellow!80!white}80 & \cellcolor{mediumyellow!84!mediumred!80!white}45 \\
			(6 & 6) & \cellcolor{mediumgreen!95!mediumyellow!80!white}90 & \cellcolor{mediumgreen!93!mediumyellow!80!white}89 & \cellcolor{mediumgreen!93!mediumyellow!80!white}89 & \cellcolor{mediumgreen!90!mediumyellow!80!white}88 & \cellcolor{mediumgreen!9!mediumyellow!80!white}55 & \cellcolor{mediumgreen!68!mediumyellow!80!white}79 & \cellcolor{mediumgreen!46!mediumyellow!80!white}70 & \cellcolor{mediumgreen!48!mediumyellow!80!white}71 & \cellcolor{mediumgreen!83!mediumyellow!80!white}85 & \cellcolor{mediumgreen!4!mediumyellow!80!white}53 \\
			\end{tabular}
	\end{center}
\end{table}

\section{Conclusion}\label{sec:mu_conclusion}
In this work we studied the throughput performance of decision rules for multiple decision problems occurring in the control of RMFS.
By analyzing a total of eight output measures for a total of 1620 RCs, we found strong correlations between these.
Most interestingly a high pile-on and a short distance traveled by the robots together almost immediately account for the success of a decision rule applied to RMFS.
Hence, we propose using these two output measures as the key tactics when designing decision strategies for RMFS that aim to achieve high throughput.
In the investigated high pressure situation further performance measures like the turnover time of pick orders were also highly correlated with the unit throughput rate, which is why we focused on the throughput itself as the main metric for a successful RMFS.\\
Furthermore, we found that varying the decision rule used for solving the Pick Order Assignment affected the unit throughput rate the most.
The average unit throughput rate was twice as high for the best decision rule as it was for the worst.
This finding indicates that system engineers and warehouse operators should pay most attention to the Pick Order Assignment decision problem.
Moreover, the unit throughput rate score ranges from 25.24\% for the worst RC assessed in phase 1 to 94.81\% for the best scoring RC.
Hence, the right combination of decision rules plays a crucial role when controlling an RMFS.
We propose that future research may assess how to scale beyond the throughput performance of the merely simple decision rules investigated in this work.
However, we observe some cross-dependencies between different strategies for the core decision problems featured in this paper, e.g., the Demand PPS rule is part of the best performing and the worst performing RC.
Thus, an integrated and realistic evaluation or validation of new decision methods for RMFS is highly important, since dependencies exist and side-effects should not be neglected.
Additionally, we found that the number of different SKUs in the system has a strong impact on the unit throughput rate.
This finding is probably due to a decrease in pile-on for a higher number of SKUs.
This effect is considerably less for larger orders, presumably because for larger pick orders pile-on tends to be higher.
Having to process return orders seems to affect the unit throughput rate more, if the pick orders are large.
Moreover, we found that the performance of the ``greedy'' benchmark consistently came close to the best ranked configurations of decision rules. \\
This paper has studied solutions to several operational problems, which lead towards promising directions for future research.
Each decision rule in this study has looked at an operational problem in isolation, but heuristics that try to integrate multiple operational problems and optimize these problems jointly could achieve substantial increases in order throughput or reductions in resources used.
Investigating rules and heuristics that increase pile-on, i.e. the number of picks per handled pod, would also be of great use to practitioners.\\
While many decision rules and parameters were varied to deliver insightful results we expect even more insight when varying the layout itself.
For example, we expect a larger impact of the PSA rule selection when facing huge layout instances.
This was not done in this work in order to keep a certain focus and to keep computational resource utilization for the conducted experiments tractable.
RMFSs are a new category of automated systems and concepts specific to RMFSs have not received much scholarly attention. 
An example would be cache zoning / priority zoning, that is the implementation of special zones near the workstations where pods are stored that will be needed 
in the near future, see also \cite{A119}. Another example would be a study of the automatic sorting of the system without explicit zones. Since pods can be relocated to another storage location each time they are transported to and from a workstation, the inventory can be sorted automatically to some degree during operations.
It is not clear at which speed automatic sorting takes place or how much performance benefits from it.
Automatic sorting is a unique feature of RMFSs, but as with so many other aspects of RMFSs, it remains to be explored.

\section*{Acknowledgements}
We would like to thank the Paderborn Center for Parallel Computing (PC$^2$) for the use of their HPC systems for conducting the experiments. Marius Merschformann is funded by the International Graduate School - Dynamic Intelligent Systems, University of Paderborn.

\bibliographystyle{plain}
\bibliography{decisionrules}

\appendix

\section{Upper bound on the unit throughput rate}\label{sec:mu_upperbound}
\begin{table}[htb]
	\setlength{\tabcolsep}{2mm}
	\small
	\caption{Times for determining the upper bound on unit throughput (all times in seconds)}
	\label{tab:mu_upper_bound_times}
	\centering
	\begin{tabular}{l|p{0.86\textwidth}}
		Symbol & Explanation \\
		\hline
		$\mParTimeOStationPickItemConstant$ & Time for picking one unit from the pod (after which the robot can be released, if no further picks are necessary) \\
		$\mParTimeOStationHandleUnitConstant$ & Time for handling one unit at a station (including picking, putting, packing, etc.) \\
		$\mUBTimeDriveInbound$ & The time for the robot to move up within the queue to the pick station's waypoint  \\
		$\mUBTimeTurnOutbound$ & The time for the robot to prepare for leaving the station (turning towards exit) \\
		$\mUBTimeDriveOutbound$ & The time for the robot to clear the station's waypoint (time to cover the minimal distance) \\
	\end{tabular}
\end{table}

In the following we introduce an upper bound for the number of units picked per hour.
This can be done by considering the constant time for picking ($\mParTimeOStationPickItemConstant$) and the constant time for handling ($\mParTimeOStationHandleUnitConstant$) a unit at a pick station.
For an overview of all necessary times see Table \ref{tab:mu_upper_bound_times}.
If a robot is queueing in the buffer of a pick station, it is assumed that it already turns the right pick face of the pod towards the side where the picker will be.
Since the robot is waiting in the queue, this happens in the best case without any additional loss of time.
During the actual pick process, the robot is occupied for $\mParTimeOStationPickItemConstant$ seconds.
After this time the robot is allowed to leave the station while the overall handling time for one unit at a station of $\mParTimeOStationHandleUnitConstant$ 
can be longer.

\begin{figure}[htb]
	\centering
	\begin{subfigure}[b]{0.48\textwidth}
		\includegraphics[width = \textwidth]{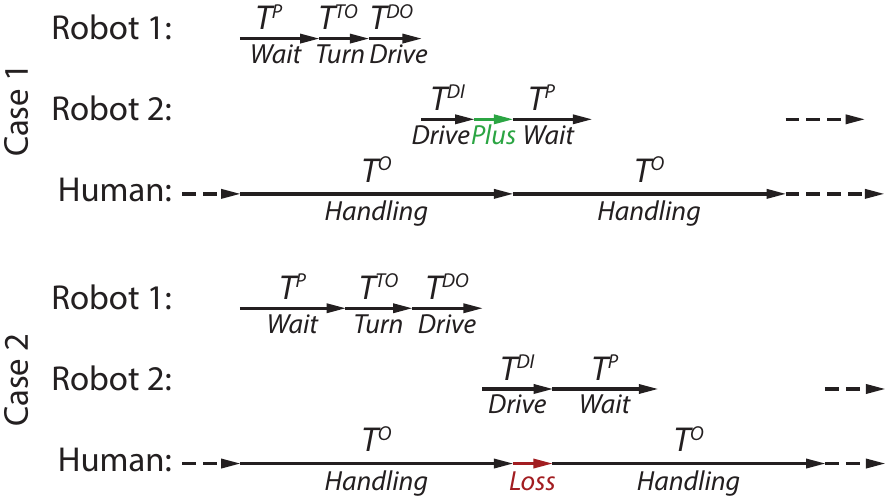}
		\caption{Sample times for both mentioned cases}
		\label{fig:mu_upperbound-sample}
	\end{subfigure}
	\,
	\begin{subfigure}[b]{0.48\textwidth}
		\includegraphics[width = \textwidth]{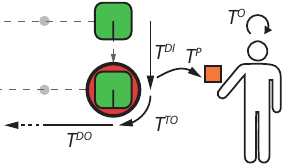}
		\caption{Times of the different steps involved in picking units}
		\label{fig:mu_upperbound-process}
	\end{subfigure}
\end{figure}

There are two cases to distinguish for obtaining a performance upper bound.
First, if the time for picking a unit from the pod plus the time for moving up the next robot (i.e. the time to turn and drive away from the station and the 
time for the next robot to approach the station from the queue area) is smaller than the overall handling time of a unit at the station, there is a surplus of 
time available on the system side and the performance is limited by the handling time of the picker (see case 1 in Figure \ref{fig:mu_upperbound-sample}). 
In the second case we face a longer time for moving up the next robot, hence, in this case we have a loss of time on the system's side and the system is limiting the throughput performance of the picker (see case 2 in Figure \ref{fig:mu_upperbound-sample}). 
For the sake of clarity we define the time for moving up the next robot in queue as $\mUBTimeMoveUpNextRobot := \mUBTimeDriveInbound + \mUBTimeTurnOutbound + 
\mUBTimeDriveOutbound$.

\begin{equation}\label{eq:upperbound}
UB :=
\begin{cases}
\left\vert \mSetOStations \right\vert \dfrac{3600}{\mParTimeOStationHandleUnitConstant} & \mParTimeOStationPickItemConstant + \mUBTimeMoveUpNextRobot \le \mParTimeOStationHandleUnitConstant \\
\left\vert \mSetOStations \right\vert \mPerformanceItemPileOn \dfrac{3600}{\mParTimeOStationPickItemConstant + \mUBTimeMoveUpNextRobot - \mParTimeOStationHandleUnitConstant + \text{IPO} \mParTimeOStationHandleUnitConstant} & \text{else}
\end{cases}
\end{equation}

Considering both cases we can determine an upper bound on the unit throughput rate, i.e. the number of units picked per hour (see Equation \ref{eq:upperbound}). 
For the first case we only need to consider the unit handling time of the picker to determine the maximum throughput rate of one station and multiply it with the overall count of pick stations $\left\vert \mSetOStations \right\vert$. 
The second case is slightly more complicated, because we also need to consider the pile-on. 
In the denominator we first calculate the loss of time seen in Figure \ref{fig:mu_upperbound-sample} and add it to the average handling time of a pod based on the estimated number of picks from it ($\text{IPO}$). 
We calculate how many pods are handled in one hour and multiply this by the $\text{IPO}$ and the number of stations overall to get the overall upper bound.\\
We recognize that this upper bound on the unit throughput rate relies on some heavy assumptions for real systems, but still propose it as a rule-of-thumb for practitioners, since it is useful for implementations where the time for moving the next robot, the handling times and the pile-on can be estimated. 
It is a natural limit of the system's performance that the system cannot exceed, even if the number of robots is more than sufficient to supply a continuous stream of pods and all rules are performing well. 
For this work, the upper bound is correct, because all mentioned times can be accurately determined or are constant and are not subject to random influence within the simulation.

\end{document}